\documentclass[preprint,aps,showpacs,nofootinbib]{revtex4}

\usepackage{amssymb}
\usepackage{bm}
\usepackage{epsfig} 
\usepackage{graphicx}

\def\bea{\begin{eqnarray}}
\def\eea{\end{eqnarray}}
\def\bec{\begin{center}}
\def\ec{\end{center}}

\def\beq{\begin{equation}}
\def\eeq{\end{equation}}

\begin{document}

\begin{center}
{\Large \bf  Anomalous $U(1)$ Mediation in
Large Volume Compactification}
\end{center}

 \vspace*{5mm} \noindent
\centerline{Chang Sub Shin} \vskip 1cm
\centerline{\em Department of Physics, Korea Advanced
Institute of Science and Technology} \centerline{\em Daejeon
305-701, Korea}
\centerline{{\em Email}: csshin@muon.kaist.ac.kr}

\vskip .9cm

\centerline{\bf Abstract} \vskip .3cm

We study the general effects of anomalous $U(1)_A$ gauge symmetry on
soft supersymmetry (SUSY) breaking terms in large volume scenario,
where the MSSM sector is localized on a small cycle whose volume is
stabilized by the $D$-term potential of the $U(1)_A$.
Since it obtains  SUSY breaking mass regardless of the detailed form
of  K\"ahler potential, the $U(1)_A$ vector superfield acts as a
messenger mediating the SUSY breaking in the moduli sector to the
MSSM sector. Then, through the loops of $U(1)_A$ vector superfield,
there arise soft masses of the order of $m_{3/2}^2/8\pi^2$ for
scalar mass squares, $m_{3/2}/(8\pi^2)^2$ for gaugino masses, and
$m_{3/2}/8\pi^2$ for $A$-paramteres. In addition, the massive
$U(1)_A$ vector superfield can have non-zero $F$ and $D$-components
through the moduli mixing in the K\"ahler potential, and this  can
result in larger soft masses depending upon the details of the
moduli mixing.  For instance, in the presence of one-loop induced
moduli mixing between
 the visible sector modulus and the large volume modulus,
 the $U(1)_A$ $D$-term provides soft scalar mass squares of the order of
 $m_{3/2}^2$.
 However, if the visible sector modulus is
mixed only with small cycle moduli, its effect on soft terms depends
on how to stabilize the small cycle moduli.


\newpage
\section{Introduction}

Moduli stabilization is one of the key steps to understand low
energy phenomenology of string theory. So far, several scenarios are
suggested and their phenomenological and cosmological consequences
are studied extensively. In particular, within the Type IIB theory,
fluxes and non-perturbative corrections to the superpotential are
considered as crucial ingredients for fixing moduli \cite{DIN,GKP}.
Based on this idea, two types of scenario are particularly
well-studied. The first is the KKLT type scenario \cite{KKLT} in which the
K\"ahler moduli are stabilized at a supersymmetric AdS  minimum by
non-perturbative correction to the superpotential and the vacuum is
uplifted to de-Sitter spacetime by additional SUSY breaking effect
such as an anti-brane. On the other hand, if the overall volume
modulus is taken to have a large vacuum value, the non-perturbative
superpotential of the large volume modulus will be negligible. Still
the large volume modulus can be stabilized at  SUSY breaking minimum
if there exists a small cycle modulus which admits non-perturbative
superpotential and has a correct sign of $\alpha'$ correction to its
K\"ahler potential. This is the so-called Large Volume Scenario
(LVS) \cite{LVS} which we will focus on in this paper.

For both the KKLT and LVS scenarios, in realistic situation, the
number of independent and sizable non-perturbative terms in the
superpotential might not be enough to stabilize all K\"ahler moduli.
As pointed out in \cite{Blumenhagen:2007sm}, when the
non-perturbative superpotential of the visible sector K\"ahler
modulus $T_v$ is generated by E3 instantons, it must be equipped
with the standard model (SM) charged matter superfields. Because the
vacuum values of the SM charged matter fields should be zero or at
most weak scale, the effect of such non-perturbative superpotential
on fixing the visible sector modulus must be negligible.
 One natural solution for fixing $T_v$ in such case is $D$-term stabilization. 
 If there exists an anomalous $U(1)$
symmetry under which $T_v$  transforms nonlinearly, the
corresponding $D$-term contains the moduli dependent FI-term which
is proportional to $\partial_{T_v}K$. If the moduli space of the
underlying string compactification admits a solution with vanishing
FI-term, which is indeed the case for many of the Type IIB string
compactifications, the $D$-term scalar potential fixes  $T_v$ near
the point with vanishing FI-term.

Stabilizing moduli in the absence of proper non-perturbative
superpotential is not just an issue of moduli stabilization, but
directly related to the pattern of soft SUSY breaking parameters in
the visible sector. For the KKLT type scenario, the soft terms in
case with anomalous $U(1)$  have been studied in
\cite{Choi-Jeong2,Choi-Jeong3}. Combining with the SUSY breaking
effects of the original KKLT type models
\cite{mirage1,mirage1a,mirage2,Choi-Jeong1}, it has been noticed
that  various patterns of soft terms can be realized. On the other
hand, for the LVS, the soft terms generated by $D$-term
stabilization have been discussed recently  in \cite{Blumenhagen et
al.,Conlon-Pedro,Choi:2010gm}.

In  \cite{Choi:2010gm}, the structure of  soft terms has been
examined for a class of LVS in the presence of one-loop induced
moduli mixing between the visible sector modulus and the large
volume modulus \cite{Conlon-Pedro}. It was shown that such moduli
mixing induces a $U(1)_A$ $D$-term of the order of $m_{3/2}^2$,
which would provide soft scalar masses of the order of the gravitino
mass $m_{3/2}$,
while the resulting gaugino masses and $A$-parameters are of the
order of $m_{3/2}/8\pi^2$. Therefore, in such set up, the gravitino
mass cannot be much larger than the (multi) TeV scale to realize
weak scale SUSY. However, it is also noticed that the specific form
of moduli mixing plays the crucial role to determine the size of
soft masses. Such mixing-dependent soft terms can be classified as
model-dependent contribution of the $U(1)_A$ mediation. Then, it is
natural to ask if there exists  any  model-independent contribution
of the $U(1)_A$ mediation, not depending on the detailed form  of
the moduli K\"ahler potential. If such contribution exists,
it would provide the lower bound of the soft masses in generic LVS
with anomalous $U(1)_A$.

The aim of this paper is to extend the previous analysis
\cite{Choi:2010gm} to more general class of LVS. We first divide the
soft terms  into the model-independent and the model-dependent parts
on the basis of how much they depend on the detailed form of moduli
mixing in the K\"ahler potential. It is shown that the $U(1)_A$
vector supermultiplet gains SUSY-breaking mass splitting regardless
of the moduli mixing, so the model-independent soft masses are
generated as a result of the loop threshold of the massive $U(1)_A$
vector supermultiplet. The resulting soft scalar masses are of the
order of $m_{3/2}/4\pi$, while gaugino masses are of the order of
$m_{3/2}/(8\pi^2)^2$ and A-terms are of the order of
$m_{3/2}/8\pi^2$. For the model-dependent contributions, as in
\cite{Choi:2010gm}, $D$-term contribution can appear due to the
moduli mixing in the K\"ahler potential. In addition to the case
studied in \cite{Choi:2010gm}, we study the case that the visible
sector K\"ahler modulus is mixed with  other small cycle K\"ahler
moduli at  tree-level, and find that its contribution can dominate
the soft terms or not, depending on how to stabilize the small cycle
K\"ahler moduli. In any case, we find that the $U(1)_A$ mediated
soft terms play an essential role to determine the spectrum of the
MSSM soft terms for models with $D$-term stabilization in LVS.

This paper is organized as follows. In section (\ref{section00}), we
review the work of \cite{Choi:2010gm}, especially focus on the
$U(1)_A$ contribution to soft scalar mass. In section(\ref{section01}), 
we show that there are other types of soft terms induced by 
$U(1)_A$, not only those discussed in \cite{Choi:2010gm}.
Section (\ref{section02}) is devoted to construct the effective
action of the light degrees of freedom by integrating out the heavy
$U(1)_A$ vector superfield, and calculate the MSSM soft terms
discussed in (\ref{section01}) more concretely. Section
(\ref{section03}) is the conclusion. Throughout the paper, 
we will limit ourselves to 4D effective SUGRA.

\section{Review of D-term stabilization with  moduli mixing}
\label{section00}

Before moving to the central part of our argument,  it is 
worth reviewing the previous work \cite{Choi:2010gm}, 
in which we studied sparticle spectrum of large volume compactification
with loop-induced moduli mixing.

In the large volume scenario (LVS), there are at least 
two types of K\"aher moduli superfields.
One type is a large volume modulus $T_b$ which determines
the overall size of a compactification volume. Another type,
$T_s$, describes the volume of a
small 4-cycle which admits non-perturbative effects to the 
superpotential. Then, ${\rm Re}T_b$ can be 
stabilized at a large vacuum value due to the competition between 
$\alpha'$ corrections suppressed by the inverse compactification volume
and the non-perturbative corrections which are exponentially suppressed. 
In the large volume limit, $\langle 
{\rm Re}T_b\rangle\gg 1$, the model is given by\footnote{ In 
most of discussion, detailed dynamics of the
string dilaton and  complex structure moduli is not important, so
we assume that they are stabilized at a supersymmetric solution
by fluxes and regarded as fixed values \cite{LVS, Blumenhagen et al.}.
The effect of  backreaction due to the K\"ahler moduli stabilization 
is also negligible. In the following, unless specified, we set the 
4D Planck scale (in the Einstein frame) $M_{\rm Pl} = 1.$}
\bea\label{model0} K &=& -3\ln  t_b + \frac{ (t_s^{3/2} 
-\xi_{\alpha'})}{ t_b^{3/2}}+{\cal O}(t_b^{-3}), \nonumber\\
W&=&W_0 + A e^{-a T_s}\eea for $t_I= T_I+ T_I^*$ $
(I=b, s)$. $\xi_{\alpha'}$ represents the leading order 
$\alpha'$ correction, $W_0$ is the flux induced constant 
superpotential,  $A$ and $a$ are constants involved in 
the non-perturbative correction to the superpotential.
In this model,  the vacuum values of $t_b$ and $t_s$ 
are fixed as \bea at_s = 2 \ln \frac{M_{\rm Pl}}{|m_{3/2}|} 
+ {\cal O}(1),\quad t_s^{3/2}=\xi_{\alpha'}\left(1+ {\cal O}
\Big(\frac{1}{at_s}\Big)\right), \eea where  
the gravitino mass, $ m_{3/2} =e^{K/2} W= 
W_0 t_b^{-3/2}(1+ {\cal O}(t_b^{-3/2}))$. 
It is straightforward to find
 \bea \label{Ftermbs} \frac{F^{T_b}}{t_b} 
&=& m_{3/2}^*\Big(1+ {\cal O}(t_b^{-3/2})\Big),\nonumber\\
\frac{F^{T_s}}{t_s}&=& \frac{m_{3/2}^*}{\ln|M_{\rm Pl}/m_{3/2}|}
\left(\frac{3}{4}+ {\cal O}\Big(\frac{1}{\ln |M_{\rm Pl}/m_{3/2}|}
\Big)\right),\eea 
where the $F$-component of a generic chiral superfield $\Phi^I$
is defined by  $F^{I} = -e^{K/2}K^{I\bar J}(D_{J}W)^*$.
An interesting feature of the LVS is a
large hierarchy among the mass scales such as
\bea
\frac{M_{\rm st}}{M_{\rm Pl}} \sim t_b^{-3/4},\quad
\frac{M_{\rm KK}}{M_{\rm Pl}}\sim t_b^{-1},\quad
\frac{M_{\rm wind}}{M_{\rm Pl}}\sim t_b^{-1/2},\quad
\frac{m_{3/2}}{M_{\rm Pl}}\sim  |W_0| t_b^{-3/2}, 
\eea  where $M_{\rm st}$ is the string scale, $M_{\rm KK}$ is
the bulk KK scale, and $M_{\rm wind}$ is the winding scale.

In order to construct a phenomenologically viable model, 
we need to specify the visible sector.  
It is noticed in \cite{Blumenhagen:2007sm}, 
however, that the MSSM cycle K\"ahler modulus $T_v$
can not have a non-perturbative superpotential 
like $T_s$ due to the chiral nature of MSSM matter.
Besides, $T_v$ cannot be identified as $T_b$, since the MSSM gauge 
couplings at high energy scale are inversely proportional to 
the vacuum value of the modulus. Therefore  in the LVS, 
the visible sector modulus is not fixed by non-perturbative and 
$\alpha'$ corrections. In such a situation, $D$-term stabilization 
can be used to fix $T_v$. Elaborating further on the issue in the framework 
of 4D effective SUGRA, we introduce the anomalous  $U(1)_A$ 
gauge symmetry and suitable gauge transformations as followings.
\bea U(1)_A\ :\quad V_A \rightarrow V_A +\Lambda_A + \Lambda_A^*,
\quad T_v \rightarrow T_v + 2\delta_{\rm GS} \Lambda_A,\quad
\Phi_i \rightarrow e^{-2q_i \Lambda_A}\Phi_i, \eea  where
$V_A$ is the vector superfield which contains the $U(1)_A$ 
gauge boson, $\Lambda_A$ is a chiral superfield parameterizing 
the $U(1)_A$ transformation on $N=1$ superspace,  $T_v$
is the visible sector K\"ahler modulus chiral superfield which 
transforms nonlinearly under the $U(1)_A$, $\delta_{\rm GS}$ 
denotes the constant associated with Green-Schwarz (GS) anomaly 
cancelation \cite{Green-Schwarz}, and finally $\Phi_i$ stands for 
generic chiral matter superfields localized on the visible sector 
4-cycle with $U(1)_A$ charge $q_i$. Since $\delta_{\rm GS}$ 
is determined by GS anomaly cancellation condition which is 
evaluated at one-loop level, generically \bea \label{delta}
\delta_{\rm GS} = {\cal O}\Big(\frac{1}{8\pi^2}\Big).\eea

Once taking into account the above symmetry, 
we can write down the gauge invariant K\"ahler 
potential and superpotential, including the visible sector fields, proposed by
\bea \label{model1}  K &=& -3\ln  t_b + \frac{ 
(t_s-\alpha_s\ln t_b)^{3/2} -\xi_{\alpha'}}{ t_b^{3/2}}
+{\cal O}(t_b^{-3})+ \frac{ (t_A-\alpha_A\ln t_b)^2 }{2  t_b^p}  
+  Z_i\Phi_i^* e^{2q_i V_A}\Phi_i+ {\cal O}(\Phi_i^4),\nonumber\\
W &=& W_0 + A e^{-a T_s}  + {\cal O}(\Phi_i^3), \eea where
$t_I= T_I+ T_I^*$ $(I=b, s, v)$, and $t_A = t_v- 2\delta_{\rm GS} V_A$
is the gauge invariant combination of the visible sector modulus,
$p$ is the modular weight which determines the $U(1)_A$ 
gauge boson mass scale (\ref{Suckelberg}), and  $\alpha_A$ 
($\alpha_s$) is the moduli mixing parameter between $t_v$ 
($t_s$) and $t_b$. 
Several assumptions were made regarding the $U(1)_A$ sector.
First, the K\"ahler potential allows the limit of vanishing FI-term,
$\partial_{T_v} K =0$. Second, there are radiative corrections
to the K\"ahler potential at one-loop level, which induce
the moduli mixing between the visible sector modulus
and the large volume modulus, i.e. $t_A\rightarrow t_A -\alpha_A \ln t_b$.
Then the mixing parameter, $\alpha_A={\cal O}(1/8\pi^2)$.

The first assumption plays a key role in achieving the $D$-term stabilization of $T_v$. 
The idea behind the $D$-term stabilization is that $T_v$ 
becomes a part of the massive $U(1)_A$ vector superfield \cite{Choi:2010gm}.
The imaginary part of the scalar component of 
$T_v$ is eaten by the $U(1)_A$ gauge boson, gaining a mass 
through the St\"uckelberg mechanism. The real part obtains 
the same mass from the $D$-term potential. The $U(1)_A$ gaugino 
and the fermionic component of $T_v$ constitute a Dirac spinor with the  
same mass as the bosons. In the supersymmetric limit, the 
mass squared of the $U(1)_A$ vector supermultiplet is given by
\bea \label{Suckelberg} M_A^2 = \left\langle \frac{g_A^2}{2}
\frac{\partial^2 K}{\partial V_A^2}\right\rangle\simeq
\left\langle2g_A^2\delta_{\rm GS}^2 \frac{\partial^2 K}{\partial T_v
\partial T_v^*}\right\rangle  =\left\langle \frac{2g_A^2 
\delta_{\rm GS}^2}{t_b^p}\right\rangle,\eea where 
$g_A$ is the $U(1)_A$ gauge coupling. As \cite{Blumenhagen et al.}, 
if $p$ is $3/2$, \bea M_A\sim M_{\rm st}/8\pi^2 \gg m_{3/2}. 
\eea Therefore we expect that the $U(1)_A$ vector supermultiplet 
is much heavier than the remaining K\"ahler moduli and matters,
which indicates that the massive $U(1)_A$ vector superfield 
is fixed  mostly by following superfield equations of motion,
\bea\label{supereq} \frac{\partial K}{\partial V_A}\simeq
-2\delta_{\rm GS}\partial_{T_v} K\simeq 0.\eea
We can find the solution of (\ref{supereq}) provided by the first 
assumption. 
Then, $T_v$ is stabilized near the point with vanishing FI-term.

The second assumption turns out to be important in determining the pattern of soft
terms. In order to account for the effect on the soft terms, 
consider the soft scalar mass squared of $\Phi_i$ given by
\bea \label{softscalar}
m_i^2 \simeq \frac{2}{3} \left\langle
V_F + \sigma V_{\rm uplift}\right\rangle
-F^{T_I}F^{T_J^*}\partial_{T_I}\partial_{T^*_J} 
\ln e^{-K/3}Z_i -q_i g_A^2 D_A \quad (I=b, s, v)\eea
at  tree-level of 4D effective SUGRA.
$V_F= e^K \left(K^{I\bar J} D_I W (D_J W)^* - 3|W|^2\right)$ 
is the $F$-term scalar potential determined by (\ref{model1}), and 
$V_{\rm uplift}$ is an additional  
uplifting potential  needed to achieve a phenomenologically 
viable de-Sitter vacuum
$V_0\approx \langle V_F +V_{\rm uplift}\rangle\simeq 0$.
$g_A^2D_A$ is the auxiliary $D$-component of the $U(1)_A$
gauge superfield $V_A$. 
The constant $\sigma$ depends on the origin of the uplifting potential.
If $V_{\rm uplift}$ originates from an anti-brane 
(or any SUSY breaking branes) stabilized at the tip of warped throat,
then $\sigma=1$. As a result, the vacuum energy contributions 
are almost cancelled and the remaining contributions are
much suppressed compared to  $\langle V_F\rangle$.
On the other hand, if $V_{\rm uplfit}$ is made by 
$F$-term uplifting, $\sigma$ is $3/2$. In such cases,
the first term in the RHS of (\ref{softscalar}) 
is of the order
\bea-\frac{1}{3}\langle V_F\rangle = {\cal O} \left(\frac{m_{3/2}^2 t_b^{-3/2}}
{\ln (M_{\rm Pl}/m_{3/2})}\right).
\eea
To evaluate the second term (modulus-mediated contribution) and 
third term ($D$-term contribution) in the RHS of (\ref{softscalar}), 
additional terms should be specified. The matter K\"ahler metric 
$Z_i$ is given by \bea \label{metric}
Z_i = \frac{{\cal Y}_i((t_A -\beta_A\ln t_b))}{t_b}\left(1+ {\cal O}\Big(
t_b^{-3/2}\Big)\right).\eea 
Here 
${\cal Y}_i((t_A-\beta_A\ln t_b))$ 
is assumed to be expanded 
about $(t_A-\beta_A\ln t_b)=0$ 
in positive powers of $(t_A-\beta_A\ln t_b)$ 
to allow the vanishing limit of $(t_A-\beta_A\ln t_b)$
as \cite{Blumenhagen et al.}
\bea
{\cal Y} _i((t_A-\beta_A\ln t_b))= {\cal Y}_i(0) + {\cal Y}_i^{(n)}(0)
(t_A-\beta_A\ln t_b)^n + {\cal Y}^{(n+1)}_i(0) (t_A-\beta_A\ln t_b)^{n+1} + \cdots,
\eea
where $n$ is the positive integer,  and ${\cal Y}_i(0),\ {\cal Y}^{(n)}_i(0),\
{\cal Y}^{(n+1)}(0),\cdots$  are constants of order one. 
 Since the matter fields in visible sector
are localized on the small 4-cycle,
$t_b$-dependence of $Z_i$ can be understood by the argument  
that the physical Yukawa couplings should not have power-dependence 
on the bulk compactification volume. Logarithmic dependence, however, 
is allowed at the one-loop level. Hence the  moduli mixing parameter  
$\beta_A$ is also of the order of $1/8\pi^2$. The auxiliary components, 
$F^{T_v}$ and $g_A^2D_A$, are  determined dominantly by the superfield 
equations of motion (\ref{supereq}),  \bea \label{supereq2}
  \partial_{T_v} K \simeq\frac{(t_v -2\delta_{\rm GS} V_A)
  -\alpha_A \ln t_b}{t_b^p}\simeq 0. \eea 
  In the Wess-Zumino gauge, the component fields are given by
 \bea \label{supereqsol}
 t_v&\simeq& \alpha_A\ln t_b \hskip3.5cm (\hskip0.05cm 0
 \hskip 0.05cm \ {\rm component}), \nonumber\\
F^{T_v} &\simeq&  \alpha_A\frac{F^{T_b}}{t_b} =\alpha_A m_{3/2}^*
\hskip1.66cm  (F\ {\rm component}),\nonumber\\
g_A^2 D_A &\simeq& \frac{\alpha_A}{\delta_{\rm GS}}
\left|\frac{F^{T_b}}{t_b}\right|^2=
  \frac{\alpha_A}{\delta_{\rm GS}} |m_{3/2}|^2 \hskip0.4cm 
  (D \ {\rm component}).
\eea
As stated above, both $\delta_{\rm GS}$ and $\alpha_A$ are generated
at the one-loop level, and hence
\bea \label{FT3D}
\delta_{\rm GS}\sim  \alpha_A = {\cal O}\Big(\frac{1}{8\pi^2}\Big)
&\rightarrow&  F^{T_v}= {\cal O}\Big(\frac{m_{3/2}}{8\pi^2}\Big)
,\quad g_A^2 D_A= {\cal O}(m_{3/2}^2).\eea
It is straightforward  to estimate the order of the scalar masses using the
auxiliary components provided by
(\ref{Ftermbs}), (\ref{supereqsol}) and (\ref{FT3D}).
We identify that the modulus-mediated contribution is 
much smaller than $m_{3/2}^2$ :
\bea
\Big|-F^{T_I}F^{T_J^*}\partial_{T_I}\partial_{T^*_J} \ln e^{-K/3}Z_i \Big|
 \lesssim 
{\cal O}\Big( (\beta_A-\alpha_A) \left|\frac{F^{T_b}}{t_b}\right|^2\Big)
={\cal O}\Big(\frac{m_{3/2}^2}{8\pi^2}\Big).
\eea
As a result, the soft scalar mass is dominated by
the $D$-term contribution as \bea \label{scalarmass} m_i^2
\simeq -\frac{q_i \alpha_A}{\delta_{\rm GS}}
m_{3/2}^2 = {\cal O}(m_{3/2}^2),\eea
and this  is  one of the  main result of our
previous work \cite{Choi:2010gm}.

We can interpret the result  (\ref{scalarmass})
from the view of effective theory constructed by
 integrating out $T_v$ and $V_A$.  In the effective theory, 
 the $U(1)_A$ gauge symmetry does not exist anymore,
hence  no $D$-term contribution as well.
 The effect of $D$-term, however, is transformed to that 
 of modulus mediation, which originates from 
 the effective K\"ahler metric of light
 matter fields $\tilde\Phi_i = e^{q_i T_v/\delta_{\rm GS}}\Phi_i$,
\bea \label{effZ}
Z_i^{\rm eff} =
\frac{{\cal Y}_i((\alpha_A-\beta_A)\ln t_b)}
{t_b^{1+ q_i\alpha_A/\delta_{\rm GS}}}\left(1+{\cal }O(t_b^{-3/2})\right).
\eea
The soft scalar mass squared is obtained as
\bea
m_i^2 \simeq -F^{T_b} F^{T_b^*} \partial_{T_b}\partial_{T_b^*}
\ln e^{-K/3} Z_i^{\rm eff}
\simeq -\frac{q_i\alpha_A}{\delta_{\rm GS}} m_{3/2}^2.
\eea

\section{Generic features of $U(1)_A$ mediation }\label{section01}
\subsection{Model-independent contribtuion }\label{section011}

To make brief summary on the previous section, 
 when the visible
sector K\"ahler modulus is stabilized near the point with
vanishing FI-term, the SUSY breaking  of the large
volume modulus can be transmitted to the MSSM sector
by the one-loop induced moduli mixing. Its effect appears
as the $D$-term contribution,  dominating soft scalar masses. 
The vanishing limit of FI-term is generic in string 
compactification \cite{Blumenhagen et al.,uranga,
Stability wall,GS-blowing-up}, but the form of 
moduli mixing among the visible sector modulus and 
other K\"ahler moduli is rather model dependent.
That is to say that we need to figure out a model independent, i.e. 
moduli mixing independent, soft term contribution 
of the $U(1)_A$.  In order to do so, 
let us suppose that there is no moduli mixing,  $\alpha_A=\beta_A=0$.
At first sight, the scalar mass squared seems to be
much suppressed compared to the gravitino mass squared, since
from (\ref{effZ}) \bea\label{nomixing0}
m_i^2 &\sim & -F^{T_I}F^{T_J^*}\partial_{T_I}\partial_{T_J^*}
\ln e^{-K/3}Z^{\rm eff}_i \nonumber\\  &=&
-F^{T_I}F^{T_J^*}\partial_{T_I}\partial_{T_J^*} \ln {\cal Y}_i(0)
\left(1+ {\cal O}(t_b^{-3/2})\right)\leq {\cal O}
\Big(m_{3/2}^2 t_b^{-3/2} \Big), \eea where $I=(b, s)$\footnote{In 
this case, we must  add soft term contributions 
from the string dilaton and uplifting potential, but it turned 
out that their corrections are less than or similar to the value 
given by (\ref{nomixing0}) \cite{Blumenhagen et al.}.}. 
We claim that, however, there is model-independent one-loop corrections 
to the effective K\"ahler metric of light matter fields 
to yield 
 \bea \label{one-loopZ}
 Z^{\rm eff}_{i}= Z^{\rm eff}_{i({\rm tree})}
 \Big(1 -\epsilon_{Ai}\ln t_b\Big),\eea
 where $Z_{i({\rm tree})}^{\rm eff}$ is the tree-level effective K\"ahler
metric, given by (\ref{effZ}),  and $\epsilon_{A i}$ is  the constant of  
${\cal O}(1/8\pi^2)$. Accordingly, 
the soft scalar mass is not dominated by (\ref{nomixing0})  but 
modified as follows.
\bea \label{loopsoft}
m_i^2&\simeq & -F^{T_I}F^{T_J^*}\partial_{T_I}\partial_{T_J^*}
\ln e^{-K/3}Z^{\rm eff}_{i}\nonumber\\ &\simeq &
-F^{T_b}F^{T_b^*}\partial_{T_b}
\partial_{T_b^*}\ln\Big(1 - \epsilon_{A i}\ln t_b\Big)
\simeq -\epsilon_{A i} m_{3/2}^2={\cal O}\Big(
\frac{m_{3/2}^2}{8\pi^2}\Big).\eea
Note that the values of $\alpha_A,\beta_A$ are given  by
string-loop corrections, so they are not calculable in 
4D effective SUGRA. On the other hand,
$\epsilon_{A i}$ can be computed from 
the $U(1)_A$ vector supermultiplet threshold at the 
level of effective field theory.
As mentioned in section (\ref{section00}) the massive 
$U(1)_A$ vector superfield, referred to $V_H$, obtains the
mass of (\ref{Suckelberg})  in the SUSY limit. 
In reality,  $V_H$ touchs on  the SUSY breaking
superfield $T_b$ by means of mass interaction in the K\"ahler potential,
so there is small mass splitting  among component
fields of $V_H$. Then,  $V_H$ plays the role of a messenger superfield
in the visible sector. The MSSM sparticles are communicated to the large 
volume modulus $T_b$ through the loops of $V_H$,
 and have soft masses as (\ref{loopsoft}).

To be more specific, let us fix the modular weight $p$.
In fact the $D$-term mediated soft scalar masses are not
affected by $p$, and that's why we did not care much about that in the 
previous work \cite{Choi:2010gm}. However, the 
mass spectrum of the $U(1)_A$ vector supermultiplet
is highly dependent on the value of $p$, hence $\epsilon_{A i}$ and
induced soft terms are also influenced by $p$. If the visible 
sector 4-cycle is stabilized at a geometric regime,
it is natural to fix $p$ at $3/2$ such like the K\"ahler potential
of $T_s$ in (\ref{model0}). In a singular cycle regime,
we might lose the analogy to $T_s$, but it is quite
plausible that the analogy is still valid even in that case.
So, we set \bea p=3/2. \eea
As we will see in next section, the corresponding soft scalar 
mass squared (\ref{loopsoft}) is given by \bea\label{MI}
\Delta_{\rm M.I.} m_i^2= -\frac{g_A^2q_i^2}{16\pi^2} 
|m_{3/2}|^2. \eea There are also such contributions 
for gaugino masses and $A$-parameters. We call these 
soft term contributions  ``model-independent contributions"
of the $U(1)_A$, in the sense that they are independent 
of the specific form of the moduli  K\"ahler potential.

Implication of the model-independent contributions is that
(\ref{MI}) provides the lower bound of soft scalar masses,
$|m_i|\gtrsim m_{3/2}/4\pi$, (unless the model-independent
contribution is canceled by the additional model-dependent 
string-loop correction), so that the gravitino mass 
should not exceed more that the scale of (multi) ${\rm TeV}$
if the weak scale SUSY is realized in nature. Further, since
the mass squared of (\ref{MI}) is negative for any 
nonzero $U(1)_A$ charge assignment,  
(\ref{MI}) should not dominantly contribute
to the MSSM squark and slepton masses.

\subsection{Model-dependent contribution}\label{section012}

As being noted above, the model-independent soft terms of 
the $U(1)_A$ are potentially problematic. However, 
in \cite{Choi:2010gm}, we have already argue that the 
$D$-term contribution induced by moduli mixing 
can dominate over the model-independent contributions. 
Such a non-trivial $D$-term is originated from the 
non-trivial K\"ahler potential of $T_v$. Based on the viewpoint 
of moduli mixing between the visible sector modulus and 
the other SUSY breaking moduli,
the visible sector modulus can be mixed not only with the large 
volume modulus at the one-loop level, but also with small K\"ahler 
moduli at the tree-level. Such model-dependence can be
accommodated by generalizing the model
(\ref{model1}) as follows.
 \bea \label{model11}
\frac{(t_s-\alpha_s\ln t_b)^{3/2} -\xi_{\alpha'}}{t_b^{3/2}} +
\frac{(t_A - \alpha_A\ln t_b)^2}{2t_b^{3/2}} &\rightarrow  &
\frac{\Delta K(t_{s1},\cdots t_{sn_s}, t_A, \alpha_A\ln t_b)
-\xi_{\alpha'}}{t_b^{3/2}},\nonumber\\
W_0 + A e^{- a T_s} &\rightarrow & W_0 +
\sum_{j=1}^{n_w} A_j e^{-a_j T_{sj}},
\eea
where   $ t_{sj}=T_{sj}+T_{sj}^*$ $ (j=1,\cdots, n_s)$,
$n_s$ is the number of small K\"ahler moduli, 
and $n_w$ moduli of them have 
non-perturbative terms in the superpotential.
Even though a number of K\"ahler moduli are
allowed in (\ref{model11}), all the $U(1)_A$ neutral
K\"ahler moduli would be stabilized by the SUSY breaking effect
so their masses will be around $m_{3/2}$
which is much smaller  than the mass of $V_H$. Therefore
we still make use of the superfield equation (\ref{supereq})
to evaluate $F$-term of $T_v$ and $D$-term of $V_A$ as
functions of  the light moduli $F$-terms. Then,
\bea\label{FD}
F^{T_v} &=&  -e^{K/2}K^{I\bar J}(D_{J}W)^*
\simeq \sum_{I=b, s1,\cdots sn_w}
-\left(\frac{\partial_{T_I}\partial_{T^*_v}\Delta K}{\partial_{T_v}
\partial_{T^*_v}\Delta K}\right)F^{T_I},\nonumber\\
g_A^2 D_A &=& -g_A^2\eta^I K_I \simeq  \sum_{I,J=b, s1,\cdots sn_w, v}
 \frac{1}{\delta_{\rm GS}}\left(
\frac{\partial_{T_I}\partial_{T_J^*}\partial_{T_v}\Delta K}{
\partial_{T_v}\partial_{T_v^*}\Delta K} \right)F^{T_I}F^{T_J^*},
\eea  where $\eta^I = \{\delta_{\rm GS}, - q_i \Phi_i\}$ 
for $\{ T_v, \Phi_i\}$,  $\eta^I=0$ for $T_b$, $T_{s1},\cdots ,T_{sn_w}$.  
The modulus and $D$-term mediated soft scalar masses are
determined by (\ref{softscalar}) and  (\ref{FD}) after 
stabilizing the light moduli. Depending on what types of moduli 
are mixed, the order of each contribution will be different 
and can be compared with the model-independent 
contribution.\footnote{
It is noticed that the absolute SUSY breaking scale
determined by $\parallel \hskip -0.06cm F^I \hskip -0.07cm \parallel^2$ 
$ \approx |K_{I\bar I}F^I F^{I^*}|$ should be distinguished  from 
the soft SUSY breaking mass scale 
determined by $\parallel \hskip -0.06cm\Delta_I m_i^2  \hskip -0.07cm \parallel$ 
 $\approx |  F^I F^{I^*} \partial_I \partial_{\bar I}\ln (e^{-K/3} Z_i)|$.}
Notice that for the $D$-term contribution, there is
the enhancement factor $1/\delta_{\rm GS}$ of
${\cal O}(8\pi^2)$, so there might be interesting
contributions to the soft terms. We have attempted to estimate (\ref{FD}),
and its effect on the  soft scalar masses for generic moduli mixing
by assuming that $\Delta K \sim t_{sj}\sim t_v
= {\cal O}(1)$,  $\partial_{T_I}\Delta K \sim \Delta K /t_I$ for 
$I=\{s1,\cdots, sn_w, v\}$, and $\partial_{T_b}\Delta K\sim \alpha_A\Delta K/t_b$.
Then,
\bea \frac{F^{T_v}}{t_v} &\sim & \frac{F^{T_{sj}}}{t_{sj}},\nonumber\\ 
g_A^2 D_A &\sim & \frac{\alpha_A}{\delta_{\rm GS}} 
\left|\frac{F^{T_b}}{t_b}\right|^2 
+\left( \kappa_{bj}\frac{\alpha_A}{\delta_{\rm GS} }\frac{F^{T_b}}{t_b} 
\frac{F^{T_{sj}^*}}{t_{sj}} + {\rm h.c.}\right) 
+ \frac{\kappa_{j}}{\delta_{\rm GS}}\left|\frac{F^{T_{sj}}}{t_{sj}}\right|^2, \eea
where $\kappa_{bi}$, $\kappa_{j}$ are order one.
If we take that ${\cal Y}_i= e^{-K/3}Z_i$  is also the generic 
function of $t_{sj}$, $t_v$, and $\alpha_A\ln t_b$, 
\bea
-F^I F^{J^*}\partial_I\partial_{\bar J}\ln e^{-K/3}Z_i
\sim   \lambda_b\alpha_A \left|\frac{F^{T_b}}{t_b}\right|^2 
+  \lambda_{bj}\alpha_A\left(\frac{F^{T_b}}{t_b} \frac{F^{T_{sj}^*}}{t_{sj}} 
+ {\rm h.c.}\right) +  \lambda_j\left|\frac{F^{T_{sj}}}{t_{sj}}\right|^2,
\eea
where $\lambda_b$, $\lambda_{bj}$, and $\lambda_{j}$ are order one.
In this naive estimation, soft scalar masses  seem to be dominated by 
the $D$-term contribution  whether $\alpha_A=0$ or not. However, this is not 
always true, since $\Delta K$ is not a generic function of small moduli. 
Let us consider a simple example
suggested in \cite{Blumenhagen:2007sm}\bea\label{ex1}
\Delta K &=& \Big(t_1 - \delta_{\rm GS} V_A\Big)^{3/2}
+ \sqrt{5} \Big(t_2 + \delta_{\rm GS} V_A\Big)^{3/2},\nonumber\\
W&=&W_0 + A e^{-a(T_1+ T_2)}. \eea
where $t_I= T_I+ T_I^*$ ($I=1,2$) are small cycle moduli, 
charged under the $U(1)_A$.
We change the basis of small moduli 
into the $U(1)_A$ neutral modulus $T_s= T_1+ T_2$, and
the so-called visible sector modulus $T_v= T_1- T_2$.
In this basis, (\ref{ex1}) is rewritten as
\bea\label{modelex}
\Delta K &=& \frac{1}{2\sqrt{2}}\Big(t_s + t_A\Big)^{3/2}
+ \frac{\sqrt{5}}{2\sqrt{2}}\Big(t_s- t_A\Big)^{3/2},
\nonumber\\
 W&=& W_0 + A e^{-a T_s}.
\eea
It is noticed that in the K\"ahler potential of (\ref{modelex}), 
there is no moduli mixing between 
$T_v$ and $T_b$, but nontrivial mixing between 
$T_v$ and $T_s$ exists at the tree-level.
From (\ref{supereq}) and  (\ref{FD}), we have
\bea\label{FDexample}
t_v &\simeq& 2 t_s /3  ,\quad
\frac{F^{T_v}}{t_v}\simeq \frac{F^{T_s}}{t_s}
={\cal O}\Big(\frac{m_{3/2}}
{\ln|M_{\rm Pl}/m_{3/2}|}\Big)={\cal O}\Big(\frac{m_{3/2}}{8\pi^2}
\Big),\quad
g_A^2 D_A\simeq 0.
\eea
The value of $F^{T_s}/t_s$ is given by (\ref{Ftermbs}).
The $U(1)_A$ $D$-term, induced by moduli mixing,
 is rather suppressed.
Consequently, the model-dependent
soft scalar mass squared (\ref{softscalar})
is estimated as
\bea\label{MD}
\Delta_{\rm M.D.} m_i^2\simeq  - |F^{T_v}|^2
\partial_{T_v}\partial_{T_v^*}\ln {\cal Y}_i(t_A) ={\cal O}
\left(\left|F^{T_v}/t_v\right|^2\right)=
{\cal O}\Big(\frac{m_{3/2}^2}{
(8\pi^2)^2}\Big).
\eea 
Proceeding from what has been said above, it should be concluded 
 that the model-independent contributions
(\ref{MI}) still dominate soft scalar masses,
even though the non-trivial moduli mixing exists.
In (\ref{section022}), it is shown that
the patter of  (\ref{MD}) is generic in case that all small moduli
are stabilized by non-perturbative superpotential ($n_s=n_w$),
and there is no one-loop induced moduli mixing between $T_v$ and $T_b$
($ \alpha_A=\beta_A=0$). Thus it points out
 that there should be additional soft term contribution 
from the matter sector (e.g.  gauge mediation 
which is not covered in this paper), dominating soft scalar masses.

\section{Soft SUSY breaking terms in $D$-term stabilization}\label{section02}

Up to now,  we have discussed possible types of 
soft SUSY breaking terms through the  $U(1)_A$ mediation.
In what follows, we will provide more concrete formulae
of the soft terms discussed  in section (\ref{section01}).
 Because the stabilization procedure of light fields,
and  induced soft term contributions are rather clearly
described by effective theory, we will construct the effective action 
by integrating out the massive
$U(1)_A$ vector supermultiplet. After that, 
the soft terms will be analyzed in details.

Begining from the generalized action discussed in (\ref{section012}),
there are $n_s+2$ K\"ahler moduli. Among them, one modulus
$T_b$ has the large vacuum value, and the rest of
 $n_s+1$ moduli remain small. The small moduli are classified into
one visible sector modulus $T_v$ charged under the $U(1)_A$,
$n_w$ moduli $T_{s1},\cdots, T_{s n_w}$ which have
 non-perturbative  superpotential,
and $n_s-n_w$ moduli $T_{sn_w+1},\cdots,T_{s n_s}$ 
which do not have non-perturbative terms in the superpotential.
The visible sector matter fields $\Phi_i$ are localized on a small 4-cycle
whose volume is described by $T_v$.
The holomorphic gauge kinetic functions of the $U(1)_A$ and
the MSSM gauge groups are  referred to $f_A$ and $f_a$ respectively.
Then, the K\"ahler potential, superpotential and gauge kinetic functions are given by
\bea\label{model2}
K &=& -3\ln  t_b + \frac{\Delta K( \vec t_{s}, t_A,
\alpha_A\ln t_b)-\xi_{\alpha'}}{ t_b^{3/2}}
+{\cal O}(t_b^{-3}) +
 Z_i\Phi_i^* e^{2q_i V_A}\Phi_i
+ {\cal O}(\Phi_i^4),\nonumber\\
W &=& W_0 + \sum_{j=1}^{n_w} A_j e^{-a_j T_{sj}}
+ \frac{1}{3!}\lambda_{ijk}(\vec T_s)\Phi_i\Phi_j\Phi_k
+{\cal O}(\Phi_i^4),\nonumber\\
f_{A}&=& k_{A} T_v +\gamma_{A}(\vec T_{ s}),\quad
f_{a}= k_{a} T_v +\gamma_{a}(\vec T_{s})
\eea
where
\bea t_I&=&T_I+T_I^*\hskip1.3cm  {\rm for }\quad  I=b, s1,\cdots, sn_s, v, \nonumber\\
\vec g&=& g_{1},\cdots, g_{n_s} \quad\quad {\rm for}\quad g=t_s, T_s,\nonumber\\
Z_i&=&Z_i(\vec t_{ s}, t_A, t_b)= \frac{{\cal Y}_i(\vec t_{ s}, t_A,
 \beta_A\ln t_b)}{t_b} \left(1+{\cal O}(t_b^{-3/2})\right),\eea
and $k_A, k_a$ are fixed by GS anomaly cancellation conditions,
 \bea \label{GS0}   \frac{1}{4\pi^2}\sum_i q_i^3
= k_A\delta_{\rm GS},\quad
\frac{1}{4\pi^2}\sum_i q_i {\rm Tr}(T_a^2(\Phi_i))
= k_a \delta_{\rm GS},\eea where $\delta_{\rm GS}$
is encoded in the gauge invariant combination
 $t_A=t_v-2\delta_{\rm GS} V_A$.
We follow the normalization convention of \cite{Choi:2010gm},
so that the orders of each constant are given by\bea
&&a_{1},\cdots, a_{ n_w}= {\cal O}(8\pi^2),\
\xi_{\alpha'}={\cal O}(1),\  k_{A,a}={\cal O}(1),\
\delta_{\rm GS}\sim\alpha_A\sim \beta_A
= {\cal O}\Big( \frac{1}{8\pi^2}\Big).
\eea
Because we are taking bottom-up approach, 
we can not determine the moduli dependent functions $\gamma_a(\vec T_s)$,
$\lambda_{ijk}(\vec T_s)$, and ${\cal Y}_i(\vec t_s, t_A,\beta_A\ln t_b)$
whose explicit forms are given by underlying string theory. 
Although their specific expressions are required to 
calculate the soft SUSY breaking terms, the order of 
their contributions can be estimated 
under the assumption that the functions depend on $\vec T_s$ in two ways.
One way is that the visible sector cycle is sequestered from 
the cycles whose volumes are described by $\vec T_s$,
that $\gamma_a$, $\lambda_{ijk}$, and ${\cal Y}_i$
are independent of $\vec T_s$. Another way is that 
those cycles are not sequestered from each other, so $\gamma_a$, 
$\lambda_{ijk}$ and ${\cal Y}_i$ are of the same order of
$T_{sj}$ multiplied by the derivative of them with respect to $T_{sj}$.

\subsection{Effective theory}\label{section021}

Firstly, let us decompose the $U(1)_A$ vector superfield as $V_A =
V_0 + V_H$, where $V_H$ is the heavy vector superfield, and
$V_0$ is the background superfield defined by a solution of
the superfield equation of motion,\bea \label{partial V}
\left.\frac{\partial  K}{\partial V_A}\right|_{V_A= V_0}
={\cal O}(D^2\bar D^2 V_0).
\eea Ignoring the part of the supercovariant derivatives,
the K\"ahler potential is written as \bea
K = \left. K\right|_{V_A=V_0} + \frac{1}{2}\left.
\frac{\partial^2 K}{\partial V_A^2}\right|_{V_A=V_0}
 V_H^2 +  {\cal O}(V_H^3). \eea By integrating out $V_H$
 in a supersymmetric way, we get the tree-level K\"ahler potential,
 where $V_0$ is substituted,  as well as the Coleman-Weinberg type
 K\"ahler potential  at the one-loop level \cite{Grisaru:1996ve,Brignole:2000kg}. Thus
 the effective K\"ahler potential  is  given by
\bea\label{eff K} K_{\rm eff} = \left. K\right|_{V_A= V_0} +
\frac{{\cal M}_A^2}{16\pi^2}{\rm Tr}\ln
\frac{{\cal M}_A^2}{e{\cal M}_{\rm UV}^2} +
(\textrm{two-loops}), \eea where $e=2.718...$ 
is Euler's number, ${\cal M}_A^2$ is the
mass squared superfield for the $U(1)_A$ vector superfield
\bea\label{gaugemass} {\cal M}_A^2 =\left. \frac{g_A^2}{2}
\frac{\partial^2 K}{\partial V_A^2} \right|_{V_A=V_0} \eea
and ${\cal M}_{\rm UV}^2$ is the cut-off superfield which
will be specified later. To proceed further, 
let us define several superfields as  functions of $V_A$,
\bea \label{super scale}
\xi_{\rm FI}(V_A) &=& \frac{\delta_{\rm GS}
\Delta K'(\vec t_{ s}, t_A, \alpha_A\ln t_b)}{t_b^{3/2}},\quad
{\cal M}_{{\rm mat}_D}^2(V_A) = q_i\hat Z_i
\Phi_i^* e^{2q_i V_A}\Phi_i, \nonumber\\
{\cal M}_{\rm GS}^{2}(V_A) &=& \frac{\delta_{\rm GS}^2
\Delta K''(\vec t_{ s}, t_A, \alpha_A\ln t_b)}{ t_b
^{3/2}},\quad {\cal M}_{\rm mat}^2(V_A)
= q_i^2\tilde Z_i\Phi_i^* e^{2q_i V_A}\Phi_i,\eea
where $q_i \hat Z_i = q_i Z_i -\delta_{\rm GS}
{Z_i}'$, and $q^2_i\tilde Z_i=
q_i^2 Z_i - 2\delta_{\rm GS} q_i{ Z_i}'
+\delta_{\rm GS}^2 {Z_i}''$. Here,
the primed notation denotes the partial derivative
with respect to $t_A$, i.e. $f'=\partial f/
\partial  t_A$,  $f''=\partial^2 f/\partial t_A^2$.
The subscript `mat$_D$' of ${\cal M}_{{\rm mat}_D}^2$
stands for  matter fields contribution to the $D$-term.
For the model of (\ref{model2}),
up to leading order in the volume expansion, (\ref{partial V})
is equivalent to \bea\label{partial V1}
\xi_{\rm FI}(V_0) - {\cal M}_{{\rm mat}_D}^2(V_0)
={\cal O}(D^2\bar D^2 V_0).\eea
And the mass squared superfield (\ref{gaugemass}) is given by
\bea\label{gaugemass1}
{\cal M}_A^2 = 2g_A^2\Big({\cal M}_{\rm GS}^2(V_0)
+ {\cal M}_{\rm mat}^2(V_0)\Big).\eea
Comparing with  (\ref{Suckelberg}), 
the expression (\ref{gaugemass1})
includes the contribution from the charged matter fields.
In order to realize the $D$-term stabilization of $T_v$, such matter contribution
should be small and  treated  perturbatively. Therefore
we focus on the region :   \bea \label{conditionD}
M_{\rm GS}^2 \gg M_{\rm mat}^2,\eea
where $ M_{\rm GS}^2 =\langle {\cal M}_{\rm GS}^2
\rangle$ and   $ M_{\rm mat}^2=\langle {\cal M}_{\rm mat}^2\rangle$.
(\ref{conditionD}) implies that the
St\"uckelberg mechanism dominantly
determines the $U(1)_A$ gauge boson mass, i.e.
$M_A^2= \langle {\cal M}_A^2\rangle \simeq
2g_A^2 M_{\rm GS}^2$\footnote{Of course, we have to show that
the matter fields are really stabilized  far below $M_{\rm GS}$.
However, this is rather model-dependent question involving
details of the matter sector. Since we concentrate on
the soft term contributions from the moduli sector, matter field
stabilization will not be covered in this paper.}.
If there happens to be no cancellation among 
various terms inside
${\cal M}_{{\rm mat}_D}^2$,  the orders of  $\langle
{\cal M}_{{\rm mat}_D}^2\rangle $,  $\langle {\cal M}_{\rm mat}^2
\rangle$ and $M_{\rm mat}^2$ will be the same.
In this case, we can solve (\ref{partial V1}) perturbatively by
decomposing $V_0$ into $v_0 + \epsilon$, where $v_0$ is
the zeroth order vector superfield which is 
determined by the moduli sector : \bea\label{vanishing FI}
\xi_{\rm FI}(v_0)=0,\eea
and $\epsilon $ is the small expansion parameter
determined by  $v_0$ as follows\bea
\epsilon =V_0 -v_0&=& \frac{1}{2}\left(\frac{\
\xi_{\rm FI}(v_0)-{\cal M}_{{\rm mat}_D}^2(v_0) }
{{\cal M}_{\rm GS}^2(v_0)+
{\cal M}_{\rm mat}^2(v_0)}\right)\Big(1 + {\cal O}(\lesssim
\epsilon)\Big)\nonumber\\
&=& -\frac{1}{2}\left(\frac{{\cal M}_{{\rm mat}_D}^2}
{{\cal M}_{\rm GS}^2}\right)\left(1 + {\cal O}
\left(\frac{{\cal M}_{\rm mat}^2}{{\cal M}_{\rm GS}^2}\right)\right). \eea
where we have omitted $v_0$ dependence in the last line for the simplicity.
Then, up to the order of $ {\cal M}_{\rm mat}^2/
{\cal M}_{\rm GS}^2$, the background superfield $V_0$ 
can be expanded by  \bea\label{V0}
 V_0=
 \frac{1}{2} \left(\frac{t_v-t^0_A}{\delta_{\rm GS}}
-\frac{{\cal  M}_{{\rm mat}_D}^2}{{\cal M}_{\rm GS}^2}\right),
\eea  where
$t^0_A=t^0_A(\vec t_{ s}, \alpha_A\ln t_b)$ is the solution of
\bea\label{t0A}
\Delta K'(\vec t_s, t^0_A, \alpha_A\ln t_b)=0.\eea
We assume that the solution actually exists inside or 
on the boundary of K\"ahler cone.
After integrating out $V_H$, the light degrees of
freedom can be described in the  $U(1)_A$ gauge
invariant field basis.  With the matter
field redefinition $\Phi_i \rightarrow  e^{- q_i T_v/\delta_{\rm GS}}\Phi_i $,
the one-loop effective K\"ahler potential (\ref{eff K})
is given by \bea \label{eff kahler}
K_{\rm eff}&=& - 3\ln t_b + \frac{\Delta K_{\rm eff}-
\xi_{\alpha'}}{t_b^{3/2}} + {\cal O}(1/t_b^3) \nonumber\\
&&+\ Z_i^{\rm eff}\Phi_i^*\Phi_i - \frac{(q_i\hat Z_i^{\rm eff}
\Phi_i^*\Phi_i)^2}{2{\cal M}_{\rm GS}^2} \left( 1
+ {\cal O}\left(\frac{g_A^2}{8\pi^2},  \frac{{\cal M}_{\rm mat}^2}
{{\cal M}_{\rm GS}^2}\right)\right) + {\cal O}(\Phi_i^4).
\eea The one-loop correction gives
\bea \label{modmetrc}\Delta K_{\rm eff}
&=&\Delta K(\vec t_s, t^0_A, \alpha_A\ln t_b) +
\frac{g_A^2 \delta_{\rm GS}^2 }{8\pi^2}
\Delta K''(\vec t_s, t^0_A, \alpha_A\ln t_b)
\ln\frac{2g_A^2{\cal M}_{\rm GS}^2}
{e {\cal M}_{\rm UV}^2}, \nonumber\\
Z_i^{\rm eff} &=& e^{-q_i t^0_A/\delta_{\rm GS}}\left(
Z_i(\vec t_{ s}, t^0_A, t_b) + \frac{g_A^2 q_i^2
\tilde Z_i(\vec t_{ s}, t^0_A, t_b)}{8\pi^2}
\ln\frac{2g_A^2{\cal M}_{\rm GS}^2}
{{\cal M}_{\rm UV}^2}\right),\eea
where 
\bea &&q_i\hat Z_i^{\rm eff} = e^{-q_i t^0_A/\delta_{\rm GS}}\Big(
q_i Z_i(\vec t_s, t^0_A, t_b) -\delta_{\rm GS}{Z_i}'
(\vec t_s, t^0_A, t_b) \Big),\nonumber\\
&&  q^2_i\tilde Z_i(\vec t_s, t^0_A, t_b) =
q_i^2 Z_i(\vec t_s, t^0_A, t_b) - 2\delta_{\rm GS} q_i{ Z_i}'(\vec t_s, t^0_A, t_b)
+\delta_{\rm GS}^2 {Z_i}''(\vec t_s, t^0_A, t_b).
\eea The effective superpotential and gauge kinetic functions are
\bea\label{eff super}
W_{\rm eff} &=& W_0 +
\sum_{j=1}^{n_w} A_j e^{-a_j T_{sj}}  +
\frac{1}{3!}\lambda_{ijk}(\vec T_s)\Phi_i\Phi_j\Phi_k
+{\cal O}(\Phi_i^4),\nonumber\\
f^{\rm eff}_{a}&=& \gamma_{a}(\vec T_{s}),\eea where
$f_a^{\rm eff}$ is obtained by adding the anomalous pieces
generated from the matter fields redefinition. 

Let us illustrate the form of $Z_i^{\rm eff}$ in more detail 
in order to clarify the model-independent contribution.
We consider the model of (\ref{model1}). 
$\Delta K$ and $Z_i$ are  given by 
\bea \Delta K(\vec t_s, t_A, \alpha_A\ln t_b) &=&  
(t_s-\alpha_s\ln t_b)^{3/2} + (t_A-\alpha_A\ln t_b)^2/2, \nonumber\\
Z_i(t_s, t_A, t_b) &=& \frac{1}{t_b}\left({\cal Y}_i(0) + {\cal Y}_i^{(n)}(0) 
(t_A-\alpha_A\ln t_b)^n+ {\cal O}\Big((t_A-\alpha_A\ln t_b)^{n+1}\Big)\right),
\eea where $n$ is the positive integer.
In this example, we set $\alpha_A=\beta_A$ which implies 
that  $\ln t_b$-dependence of the matter K\"ahler metric comes only from
moduli redefinition. (\ref{super scale}) and (\ref{t0A}) yield
${\cal M}_{\rm GS}^2=\delta_{\rm GS}^2 t_b^{-3/2}$ and 
$t^0_A=\alpha_A\ln t_b$, respectively. As a result, 
\bea \label{Zeff}
Z^{\rm eff}_i &=& 
\frac{{\cal Y}_i(0)}{t_b^{1+ q_i \alpha_A/\delta_{\rm GS}}}
\left(1 + \frac{g_A^2 q_i^2}{8\pi^2}\ln\frac{2g_A^2 \delta_{\rm GS}^2}
{{\cal M}_{\rm UV}^2 t_b^{3/2}} \right).\eea
The effect of moduli redefinition is encoded in the prefactor 
of the RHS of (\ref{Zeff}). 

Now, we should specify the cut-off 
superfield ${\cal M}_{\rm UV}^2$
in order to determine the model-independent contribution.
We might choose ${\cal M}_{\rm UV}^2$ so that
$\langle {\cal M}_{\rm UV}\rangle \sim 
M_{\rm string}$. 
However, the cut-off scale as a ``superfield" is rather subtle 
from the 4D effective field theory point of view. 
There is no reason to take ${\cal M}_{\rm UV}^2 \sim  t_b^{-3/2}$.
By performing component calculation in appendix (\ref{append1}),
we find that ${\cal M}_{\rm UV}^2\sim t_b^{-1}$ 
is correct choice regardless of moduli redefinition.
In (\ref{append1}),  it is identified that the
tree-level mass splitting of the $U(1)_A$ vector supermultiplet
is given by  (\ref{massB}), (\ref{massF}). Through
the vector supermultiplet loops, the matter sector soft terms are generated.
In the superfield approach, this is equivalently
related to the mismatch between
${\cal M}_{\rm GS}^2$ and ${\cal M}_{\rm UV}^2$, so that
at the one-loop level the matter K\"ahler metric depends on
the large volume modulus $T_b$ as follows
\bea\label{simmetric}
Z_i^{\rm eff}= Z_{i({\rm tree})}^{\rm eff}\left(1 + \frac{g_A^2 q_i^2}{8\pi^2}
\ln\frac{{\cal M}_{\rm GS}^2}{{\cal M}_{\rm UV}^2}\right)
\simeq Z_{i({\rm tree})}^{\rm eff}\left(1- \frac{g_A^2 q_i^2}{16\pi^2}\ln t_b\right).
\eea
where $Z^{\rm eff}_{i(\rm tree)}={\cal Y}_i(0)
t_b^{-(1+ q_i \alpha_A/\delta_{\rm GS})}$
is given by (\ref{one-loopZ}) and (\ref{Zeff}).
Induced soft terms are the same as
those evaluated in the component Lagrangian.

We might infer the UV scale $\langle {\cal M}_{\rm UV}\rangle = M_{\rm UV}$
from a running gauge
coupling constant. The Kaplunovsky-Louis formula for
the physical gauge coupling \cite{gauge-coupling}
 is given by \bea
\frac{1}{g_a^2(\mu)} = {\rm Re}(f_a)
+ \frac{b_a}{16\pi^2}\ln \frac{e^{K/3}M_{\rm Pl}^2}{\mu^2}
-\frac{{\rm Tr}(T_a^2(\Phi_i))}{8\pi^2}\ln e^{-K/3}Z_i(\mu)
+\frac{{\rm Tr}(T^2_a(G))}{8\pi^2} \ln g_a^{-2}(\mu),
\nonumber\\
\eea
where $b_a = \sum_r {\rm Tr}(T_a^2(\Phi_i))-{\rm Tr}(T_a(G))$.
The combination of $e^{-K/3}Z_i(\mu)\simeq{\cal Y}_i $
is nearly independent of $t_b$ at leading order.  Thus in the large
volume limit, the effective UV scale at which the gauge
couplings start to  run  is neither the string scale
$M_{\rm st}\sim M_{\rm Pl}t_b^{-3/4}$ nor the Planck scale,
but rather the winding scale $e^{K/6}M_{\rm Pl}\sim M_{\rm Pl} 
t_b^{-1/2}$ \cite{Conlon:2009qa}.
If the holomorphic gauge kinetic functions $f_a$ are universal
as a consequence of GUT, the phyiscal gauge couplings
seem to be unified at this scale. On the one hand, when the Planck scale is
introduced in the superpotential as the natural suppression scale
of higher dimensional operators, the physical suppression
scale is not the Planck scale, but scales to $e^{K/6}M_{\rm Pl}$
due to the canonical normalization of the matter fields.
With these considerations, we naturally expect the ``effective" cut-off
of the visible sector is $M_{\rm UV}= e^{K/6}M_{\rm Pl}\simeq t_b^{-1/2}$,
and the corresponding cut-off superfield,
\bea \label{super cut off} {\cal M}_{\rm UV}^2 =
e^{K/3} M_{\rm Pl}^2 = t_b^{-1}\left(1+{\cal O}(t_b^{-3/2})\right). \eea
Again, 
we should address that the moduli-superfields dependence of ${\cal M}_{\rm UV}^2$
is not explicitly determined by the real cut-off scale of the effective SUGRA
given by underlying string theory. 
In the language of component calculations, the scalar mass contribution from 
loops of the $U(1)_A$ vector supermultiplet is the threshold correction generated at 
the scale of the $U(1)_A$ vector boson mass. 
Therefore, the real cut-off of the theory does not play the crucial role 
to determine the value of model-independent scalar masses as long as 
the cut-off scale is sufficiently bigger than the scale of the $U(1)_A$ vector 
boson mass\footnote{As we can see in (\ref{append1}),  
the additional quartic term which depends
on the cut-off scale can emerge at one-loop level, but this term 
does not contribute to the scalar mass.}.
Such UV-insensitivity of the correction is the same with that of 
gauge mediation, where 
the soft masses are generated at the scale of messenger mass and 
the UV cut-off scale of the theory is not important. 
However, we also notice that depending on the cut-off scale
of the effective theory,
there might be additional string-loop correction which cancels
the model-independent contribution obtained by (\ref{simmetric}).
If cancellation is exact at leading order,  the soft scalar masses can 
be further suppressed compared to the gravitino mass. 
Therefore evaluating such string contributions is very important.
Since the calculation is beyond the scope of this paper, we just 
mention its importance.

It is straightforward to stabilize light scalar fields
by minimizing the effective SUGRA potential constructed from
(\ref{eff kahler}) and (\ref{eff super}),
\bea V^{\rm eff}_F= e^{K_{\rm eff}}
\left(K^{I\bar J}_{\rm eff} D_I W_{\rm eff}
D_{\bar J}W^*_{\rm eff} - 3 |W_{\rm eff}|^2\right). \eea
The light K\"ahler moduli will be stabilized in the
same manner as usual LVS models.
The one-loop correction to $\Delta K_{\rm mod}^{\rm eff}$
is actually three-loop suppressed, since $\delta_{\rm GS}
={\cal O}(1/8\pi^2)$, so negligible for moduli stabilization.
The moduli $F$-components
are mostly determined by vacuum values of the scalar moduli,
where the $F$-term is defined as
\bea F^I = - e^{K_{\rm eff}/2} K^{I\bar J}_{\rm eff}
D_{\bar J} W_{\rm eff}^*.\eea
Those $F$-terms play the role of  
the SUSY breaking sources for the MSSM sector.

\subsection{Soft SUSY breaking terms}\label{section022}

We are now ready to calculate the MSSM soft terms
induced by moduli stabilization with the $U(1)_A$.
For the action described by (\ref{eff kahler}) and (\ref{eff super}),
 the MSSM soft terms take the form \bea
{\cal L}_{\rm soft}=-\frac{1}{2}M_a \lambda^a\lambda^a -
\frac{1}{2}m_i^2 |\phi_i|^2 - \frac{1}{3!}A_{ijk}y_{ijk}
\phi_i\phi_j\phi_k + h.c. ,\eea
where  $\lambda_a$ and $\phi_i$ are canonically
normalized gauginos and scalar components of $\Phi_i$
respectively, $y_{ijk}$ denote the canonically normalized
Yukawa couplings, \bea y_{ijk}=\frac{\lambda_{ijk}(\vec T_s)}{\sqrt{e^{-K_0}
Z_i^{\rm eff}Z_j^{\rm eff}Z_k^{\rm eff}}}, \eea
and the soft SUSY breaking parameters at a scale just
below $M_A$ are given by \cite{
moduli-mediation,d-term-mediation1,
d-term-mediation2,gaugino-loop,anomaly-mediation}
\bea\label{soft term1} \frac{M_a}{g_a^2} &=& \frac{1}{2}
F^I\partial_I f_a^{\rm eff} - \frac{1}{8\pi^2}
\sum_i{\rm Tr}(T_a^2(\Phi_i))F^I\partial_I\ln
\left(e^{-K_0/3}Z_i^{\rm eff}\right)+\frac{\Delta_{\rm anom.} M_a}{g_a^2},\nonumber\\
m_i^2 &=& \frac{2}{3} V^{\rm eff}_F - F^I F^{J^*} \partial_I
\partial_{\bar J}\ln \left(e^{-K_0/3}Z^{\rm eff}_i \right)
+ \Delta_{\rm anom.} m_i^2, \nonumber\\
A_{ijk}&=& - F^I\partial_I\ln\left(\frac{\lambda_{ijk}(\vec T_s)}
{e^{-K_0}Z^{\rm eff}_i Z^{\rm eff}_j Z^{\rm eff}_k}\right)
+ \Delta_{\rm anom.}A_{ijk},\eea
where $K_0= K_{\rm eff}|_{\Phi_i=0}$ and $I= T_b$, $\vec T_s$.
 The additional contribution to the soft
parameters, denoted as $\Delta_{\rm anom.}M_a$,
$\Delta_{\rm anom.}m_i^2$ and  $\Delta_{\rm anom.}A_{ijk}$,
represents the anomaly mediation \cite{anomaly-mediation}
in which the induced parameters are proportional to
\bea \left| m^*_{3/2} + \frac{1}{3} K_IF^I\right| \leq {\cal O}
\Big(m_{3/2} t_b^{-3/2}\Big) \eea
multiplied by additional loop suppression factors.
Those contributions are strongly suppressed
with respect to the prior contributions in the large
volume limit due to the no-scale property
of the leading order scalar potential, so we
neglect its effect from now on. By substituting (\ref{modmetrc}) 
to (\ref{soft term1}), and expanding  in powers of 
$1/t_b$, $g_A^2/8\pi^2$, and  $\delta_{\rm GS}$,
the leading order contributions are obtained as follows.\bea\label{soft term2}
\frac{M_a}{g_a^2} &\simeq& -
\frac{g_A^2q_a^2}{(8\pi^2)^2} m^*_{3/2}
+ \frac{1}{2} \sum_{I=T_b,\vec T_s}F^{I}\partial_{I}
\Big(\gamma_a+ k_a t^0_A\Big)
\nonumber\\
&&-\   \frac{1}{8\pi^2}\sum_i {\rm Tr}(T_a^2(\Phi_i))
\sum_{I=T_b,\vec T_s}F^{I}\partial_{I}\left(\ln {\cal Y}_i
+ \frac{g_A^2 q_i^2}{8\pi^2}\ln g_A^2\Delta K''\right),
\nonumber\\
m_i^2&\simeq & -\frac{g_A^2 q_i^2}{16\pi^2} |m_{3/2}|^2
+ \sum_{{I,J}=T_b,\vec T_s} F^{I}F^{\bar J}
\partial_{I}\partial_{\bar J}\left(\frac{q_i}{\delta_{\rm GS}} t^0_A
-\ln {\cal Y}_i - \frac{g_A^2 q_i^2}{8\pi^2}\ln g_A^2 \Delta K''\right),\\
A_{ijk}&\simeq & \frac{g_A^2(q_i^2+q_j^2+q_k^2)}{16\pi^2} m_{3/2}^*
-\sum_{I=T_b,\vec T_s}F^I\partial_I\left(\ln\frac{\lambda_{ijk}}
{{\cal Y}_i{\cal Y}_j{\cal Y}_k}
- \frac{g_A^2(q_i^2 + q_j^2+ q_k^2)}{8\pi^2}\ln g_A^2 \Delta K''\right).
\nonumber
  \eea
where \bea
&&  q_a^2=\sum_iq_i^2 {\rm Tr}(T_a^2(\Phi_i)),\quad
\gamma_a=\gamma_a(\vec T_s),\quad
\lambda_{ijk}=\lambda_{ijk}(\vec T_s),\quad
t^0_A=t^0_A(\vec t_s, \alpha_A\ln t_b),\nonumber\\
&&{\cal Y}_i= e^{-K_0/3}Z_i(\vec t_s, t^0_A, t_b)
={\cal Y}_i(\vec t_s, t^0_A, \beta_A\ln t_b),\quad
\Delta K''=\Delta K''(t_a, t^0_A, \alpha_A\ln t_b).\eea
For each soft parameters in (\ref{soft term2}),
first terms of the RHS  are
the model-independent contributions induced by the $U(1)_A$
threshold correction. These contributions are estimated as
\bea
\Delta_{\rm M.I.}M_a &=&{\cal O}\Big(\frac{m_{3/2}}{(8\pi^2)^2}\Big)
,\quad
\Delta_{\rm M.I.}m_i^2 = {\cal O}\Big(\frac{m_{3/2}^2}{8\pi^2}\Big)
,\quad
\Delta_{\rm M. I.} A_{ijk}= {\cal O}\Big(\frac{m_{3/2}}{8\pi^2}\Big).
\eea
On the other hand, the remaining contributions are determined after specifying the forms
of $\gamma_a$, $\lambda_{ijk}$, ${\cal Y}_i$, $\Delta K''$
and $t^0_A$.  The soft terms which depend on $t^0_A$
are easily  understood  from (\ref{V0}) :
\bea
F^{T_v}&=& \sum_{I=T_b, \vec T_s}F^I\partial_I\left( t^0_A +
\delta_{\rm GS}\frac{{\cal M}_{{\rm mat}_D}^2}{{\cal M}_{\rm GS}^2}\right),
\nonumber\\
-q_i g_A^2 D_A &=& \sum_{I,J=T_b,\vec T_s}
F^IF^{\bar J}\partial_I\partial_{\bar J} \left(\frac{q_i }{\delta_{\rm GS}} t^0_A
+q_i\frac{{\cal M}_{{\rm mat}_D}^2}{{\cal M}_{\rm GS}^2}\right).
\eea
In the perspective of UV theory, they are identified as
the modulus and  $D$-term mediated soft masses
induced by  moduli mixing.
In (\ref{soft term2}), the contributions 
from the SUSY breaking of matter fields 
are not included, because we focus on
the soft terms generated from the moduli sector
at a energy scale just below $M_A$.
In the effective theory, matter contributions come from the
higher dimensional operator of (\ref{eff kahler}),
and can be included consistently.
Their contributions should be critical
in the case that the effect of moduli mixing is suppressed.

In order to estimate model-dependent contributions,
let us look at the following cases. First, 
consider the case when there is no moduli mixing and the visible sector is
sequestered from the other moduli sector. Then,
$\gamma_a$, $\lambda_{ijk}$, ${\cal Y}_i$, $\Delta K''$
and $t^0_A$ are  independent of $T_b$, $\vec T_s$. As a result,
 \bea
 \Delta^{(1)}_{\rm M.D.}M_a= {\cal O}(m_{3/2} t_b^{-3/2}),\quad
 \Delta^{(1)}_{\rm M.D.}m_i^2\leq{\cal O}(m^2_{3/2} t_b^{-3/2}),\quad
\Delta^{(1)}_{\rm M.D.}A_{ijk}= {\cal O}(m_{3/2} t_b^{-3/2}).\eea
The model-independent
contribution dominates overall soft terms. The second case is that
the visible sector is still sequestered from the small moduli sector,
but the one-loop induced moduli mixing between
the visible sector modulus and the large volume modulus 
\cite{Conlon-Pedro,Choi:2010gm} gives rise to
$t^0_A\simeq \alpha_A\ln t_b$, ${\cal Y}_i
\simeq{\cal Y}_i((\alpha_A-\beta_A)\ln t_b)$. Then, 
 gaugino and sfermion masses are  dominated by 
the model-dependent contribution :
\bea
\Delta^{(2)}_{\rm M.D.} M_a&\simeq &\frac{g_A^2}{2} F^{T_b}\partial_{T_b}
\Big(k_a t^0_A\Big)\simeq \frac{g_A^2k_a\alpha_A}{2}\frac{F^{T_b}}{t_b}
=\frac{g_A^2k_a\alpha_A}{2}m_{3/2}^*, \nonumber\\
\Delta^{(2)}_{\rm M.D.} m_i^2 &\simeq &  \frac{q_i}{\delta_{\rm GS}} F^{T_b}
F^{T_b^*}\partial_{T_b}\partial_{T_b^*}t^0_A \simeq 
-\frac{q_i\alpha_A}{\delta_{\rm GS}} |m_{3/2}|^2,
\eea
whereas the model-dependent contribution to $A$-terms 
might be comparable with the model-independent contribution :
\bea
\Delta^{(2)}_{\rm M.D.} 
A_{ijk}&\simeq & - F^{T_b}\partial_{T_b} \ln\frac{\lambda_{ijk}}
{{\cal Y}_i{\cal Y}_j{\cal Y}_k}\nonumber\\ &\simeq& 
(\alpha_A-\beta_A) m_{3/2}^* \Big(\partial_t\ln {\cal Y}_i(t){\cal Y}_j(t)
{\cal Y}_k(t)\Big)_{t=(\alpha_A-\beta_A)\ln t_b}. \eea
The resulting model-dependent contributions are estimated as
\bea\label{modelsecond}
\Delta^{(2)}_{\rm M.D.} M_a ={\cal O}\Big(\frac{m_{3/2}}{8\pi^2}\Big),\quad
\Delta^{(2)}_{\rm M.D.} m_i^2 ={\cal O}( m_{3/2}^2),\quad
\Delta^{(2)}_{\rm M.D.} A_{ijk}= {\cal O}\Big(\frac{m_{3/2}}{8\pi^2}\Big).
\eea
The third case is that there is no one-loop induced  moduli mixing with the large volume
modulus
($\alpha_A=\beta_A=0$), but the visible sector is not sequestered 
from the small moduli sector. Hence $\{\gamma_a$, $\lambda_{ijk}\}$ 
and $\{{\cal Y}_i$, $\Delta K''$,  $t^0_A\}$ are generic functions of 
$\vec T_s$ and $\vec t_s$ respectively. Most controllable situation 
is that $n_s=n_w$, i.e. the number of small moduli is equal to the 
number of non-perturbatively generated terms in the superpotential. 
In appendix (\ref{append2}), we show that  due to the
no-scale property of the K\"ahler potential,   the leading order 
$F^{T_{sj}}/t_{sj}$  are universal as
\bea \label{smallmod}
\frac{F^{T_{sj}}}{t_{sj}}= \frac{m_{3/2}^*}{\ln|M_{\rm Pl}/m_{3/2}|}
\left(\frac{3}{4} +{\cal O}\Big(\frac{1}{\ln|M_{\rm Pl}/m_{3/2}|}\Big)\right)
={\cal O}\Big(\frac{m_{3/2}}{8\pi^2}\Big)\quad \textrm{for}\ j=1,\cdots, n_s=n_w.
\eea
At first sight, the model-dependent ($D$-term) contribution to 
the soft scalar mass seems to be comparable with the 
model-independent contribution by following estimation. \bea\label{naive}
\Delta_{\rm M. D.} m_i^2 \simeq\frac{q_i}{\delta_{\rm GS}}
\sum_{I,J=\vec T_s}F^IF^{\bar J}\partial_I\partial_{\bar J} t^0_A(\vec t_s)
={\cal O}\Big(\frac{1}{\delta_{\rm GS}}\left|\frac{F^{T_s}}{t_s}\right|^2\Big)
={\cal O}\Big(\frac{m_{3/2}^2}{8\pi^2}\Big),
\eea
where  $1/\delta_{\rm GS}={\cal O}(8\pi^2)$,
$t_{sj}t_{sk}\partial_{t_{sj}}\partial_{t_{sk}} t^0_A\sim
t^0_A= {\cal O}(1)$ according to our normalization convention.
However, this is not easily achieved.  The no-scale property of 
the tree-level K\"ahler potential imply 
that $\Delta K(\lambda \vec t_s,\lambda t_A) \approx
\lambda^{3/2} \Delta K(\vec t_s, t_A)$. So,
the solution of (\ref{t0A}) ($\Delta K'(\vec t_s, t^0_A)=0$) 
also scales as  $t^0_A(\lambda \vec t_s) 
 \approx\lambda t^0_A(\vec t_s)$. Then, the leading order 
 contributions of (\ref{naive}) cancel out :
\bea \sum_{I,J=\vec T_s}
F^IF^{\bar J}\partial_I\partial_{\bar J} t^0_A(\vec t_s) \simeq
\left|\frac{F^{T_s}}{t_s}\right|^2\sum_{t_{sj},t_{sk}}
\Big(t_{sj}t_{sk}\partial_{t_{sj}}
\partial_{t_{sk}} t^0_A\Big)  = 0
\eea thanks to the universality of  $F^{T_s}/t_s$
and the scaling behavior of $t^0_A$  at leading order. Consequently, 
the $D$-term contribution is at most of the same order as
the model-dependent modulus mediated soft term. Thus,
\bea
\Delta^{(3)}_{\rm M.D.}M_a= {\cal O}\Big(\frac{m_{3/2}}{8\pi^2}\Big),
\quad \Delta^{(3)}_{\rm M.D.}m_i ^2= {\cal O}\Big(\frac{m_{3/2}^2}{(8\pi^2)^2}\Big),
\quad \Delta^{(3)}_{\rm M.D.}A_{ijk}={\cal O}\Big(\frac{m_{3/2}}{8\pi^2}\Big).
\eea
For gaugino masses and $A$-terms, 
the model-dependent contributions are of the same order as 
those of the second case, 
\bea 
\Delta^{(3)}_{\rm M.D.}M_a &\simeq& -
\frac{g_A^2}{2} \sum_{j}F^{T_{sj}}\partial_{T_{sj}}
\Big(\gamma_a(\vec T_s) +k_a t^0_A(\vec t_s)\Big)\nonumber\\
&\simeq & -\sum_{j}\frac{F^{T_{sj}}\partial_{T_{sj}}
\Big(\gamma_a(\vec T_s) +k_a t^0_A(\vec t_s)\Big)}
{2{\rm Re}(\gamma_a(\vec T_s)) +k_a t^0_A(\vec t_s)}
\sim \frac{F^{T_{sj}}}{t_{sj}} = {\cal O}
\left(\frac{m_{3/2}}{\ln (M_{\rm Pl}/m_{3/2})}\right), \nonumber\\
\Delta^{(3)}_{\rm M.D.}A_{ijk} &\simeq & -
\sum_{j} F^{T_{sj}}\partial_{T_{sj}} \ln\frac{\lambda_{ijk}}
{{\cal Y}_i{\cal Y}_j{\cal Y}_k} \sim \frac{F^{T_{sj}}}{t_{sj}} 
={\cal O}\left(\frac{m_{3/2}}{\ln (M_{\rm Pl}/m_{3/2})} \right),\eea 
where we assume that $\gamma_a(\vec T_s)$, ${\cal Y}_i(\vec t_s, 
t^0_A(\vec t_s)), \lambda_{ijk}(\vec T_s) $ 
are generic functions of $\vec T_s$, in the sense that
$t_{sj}\partial_{T_{sj}}(\gamma_a(\vec T_s) + k_a t^0_A(\vec t_s))
\sim (\gamma_a(\vec T_s) + k_a t^0_A(\vec t_s)) $, 
$ t_{sj}\partial_{T_{sj}} {\cal Y}_i(\vec t_s, t^0_A(\vec t_s))\sim 
{\cal Y}_i(\vec t_s, t^0_A(\vec t_s))$, $ t_{sj}\partial_{T_sj} \lambda_{ijk}
(\vec T_s) \sim \lambda_{ijk}(\vec T_s)$. 
Look at the final case when  $n_s>n_w$ so that  some of
small moduli do not admit non-perturbative superpotential.
In such a case, the moduli might be stabilized via several corrections to the
K\"ahler potential which break no-scale structure
\cite{Cicoli:2007xp,Cicoli:2008va}.
Even though the situation is less controllable, we generally expect that
if the moduli are stabilized by the K\"ahler potential,
the corresponding  $F$-terms  will be of order of $m_{3/2}$.
Unlike the third case, there is no scaling property or
symmetry to suppress the $D$-term
contribution, and  $1/\delta_{\rm GS}$ enhancement effect
with respect to the ordinary modulus mediation will be realized.
Therefore, we expect
\bea
\Delta^{(4)}_{\rm M.D.}M_a= {\cal O}(m_{3/2}),
\quad \Delta^{(4)}_{\rm M.D.}m_i^2 = {\cal O}(8\pi^2 m_{3/2}^2),
\quad \Delta^{(4)}_{\rm M.D.}A_{ijk}={\cal O}(m_{3/2}),
\eea
and these contributions dominate overall soft terms.

\section{Conclusion}\label{section03}

To conclude, the central to this paper has been the study of 
soft term structure of the
LVS models in which the visible sector K\"ahler modulus
is dominantly stabilized by the $D$-term potential of the
anomalous $U(1)_A$ gauge symmetry. 
This analysis has led to the following observations : 
Regardless of the detailed form of the K\"ahler potential,
there are unavoidable soft term contributions coming from 
the $U(1)_A$ vector supermultiplet threshold correction.
These model-independent contributions are of the 
order of $m_{3/2}/4\pi$ for soft scalar masses,
$m_{3/2}/(8\pi^2)^2$ for gaugino masses, and
$m_{3/2}/8\pi^2$ for $A$-parameters. However, 
the corresponding soft scalar mass squares are negative
for  any non-zero $U(1)_A$ charge assignment.
In order to prevent charge and color breaking of the MSSM sector,
the additional model-dependent contributions must be needed.
We get such contributions from the moduli sector.
As studied in \cite{Choi:2010gm}, the moduli mixing between 
the visible sector modulus and the large volume modulus 
in the K\"ahler potential provides sfermion masses of 
the order of $m_{3/2}$. But, if the visible sector modulus
is mixed only with small moduli stabilized by non-perturbative
corrections to the superpotential, the corresponding model 
dependent contribution is of the order of $m_{3/2}/8\pi^2$.
In this case, we still need an additional contribution from the matter sector
to compensate for the model-independent sfermion mass squared.

An inevitable consequence of our paper  is that due to
the model-independent contribution, in order to obtain 
TeV-scale gaugino mass, the gravitino mass
$m_{3/2}\simeq  |W_0|t_b^{-3/2}$ is bounded
from the above by the scale of the order of $10^6{\rm GeV}$.
This means that if the effective UV scale of the visible
sector $M_{\rm UV}\sim M_{\rm Pl}t_b^{-1/2}$
is identified as the GUT scale $M_{\rm GUT}\sim 2
\times 10^{16}{\rm GeV}$, the flux induced
superpotential $W_0$ should be much smaller
than the value of ${\cal O}(1)$. In other words, if $W_0$ is
given by ${\cal O}(1)$ so that $m_{3/2}\sim 10^{11}{\rm GeV}$
for $M_{\rm UV}\sim M_{\rm GUT}$, we need to fine-tune the
visible sector model parameters  to get correct orders of soft terms.
So there is still a tension between the natural large volume
scenario and the idea of grand unification.
Of course, this conclusion can be wrong, if there are additional 
(model-dependent)
string-loop corrections which cancel the above model-independent 
contributions. Since the detailed calculation should be performed 
in string theory, we leave it as a further work.

In this paper, we did not discuss stabilization of 
$D$-flat directions. As a remnant of the $U(1)_A$,
there is anomalous global $U(1)_{\rm PQ}$ symmetry
for the $D$-flat directional light matter fields.
Since the $U(1)_{\rm PQ}$ should be  spontaneously broken
above $10^{9}{\rm GeV}$ by astrophysical  considerations \cite{axion0},
in \cite{Choi:2010gm}, we introduced the PQ sector which consists of the
$U(1)_A$ charged but the SM singlet matter fields.  They dominantly break the
$U(1)_{\rm PQ}$  and the QCD axion \cite{pq,axion1,axion2} is generated.
Similar approach can be made here. After that, we can estimate  the
soft SUSY breaking terms coming from the PQ sector. They might be keystones
when the moduli mixing effect is suppressed and the
model-independent contribution dominates  soft scalar masses.

\section*{Acknowledgments}

We would like to thank Jeong Han Kim,  
Kwang Sik Jeong, Hans Peter Nilles,  Fernando Quevedo,
and especially Kiwoon Choi for very helpful discussions and comments on the manuscript. 
This work is supported by the KRF Grants funded by the Korean Government (KRF- 2008-314-C00064 and KRF-2007-341-C00010) and the KOSEF Grant funded by the Korean Government (No. 2009-0080844).

\appendix

 \section{Model-independent soft scalar masses}
 \label{append1}

Starting from the K\"ahler potential and superpotential given by (\ref{model2}), 
let us try deriving the scalar mass squared (\ref{MI}) at a component level.
In order to see the effect of the $U(1)_A$ threshold correction clearly, 
we only consider a single small modulus and matter superfield, 
and take  $\Delta K$, $Z_i$ and $W$ as simple as possible. 
However, we allow the loop-induced moduli redefinition 
of the visible sector modulus as a probe for model-dependence. 
Then,\bea\label{toymodel}
K &=& - 3 \ln t_b + \frac{(t_s^{3/2} -\xi_{\alpha'}) + 
(t_A-\alpha_A\ln t_b)^2/2}{t_b^{3/2}}
+{\cal O}(1/t_b^{-3}) + \frac{\Phi_i^*e^{2q_i V_A}\Phi_i}{t_b},\nonumber\\
W&=& W_0 + A e^{-a T_s}.\eea The large volume
modulus $T_b$ and the small cycle modulus $T_s$
are stabilized in the usual manner. In this background, we can
extract the effective tree-level Lagrangian for  component fields of
$T_v, \Phi_i, V_A$. Since the
background spacetime is nearly flat due to the no-scale
structure of  large volume stabilization,
the leading order Lagrangian can be derived in flat spacetime limit.
In other words, we neglect any soft terms of the order of
$\Delta m_i^2 \sim m_{3/2}^2 t_b^{-3/2}$ and possible gravitational
effect. 

The tree-level Lagrangian for the canonically normalized
component fields is written as follows.
\bea\label{treeL}
{\cal L}_{\rm tree}&\simeq&\frac{1}{2}
\Big(t_v\square t_v + \varphi_v\square\varphi_v\Big)
+ \phi_i^*\square\phi_i + i \Big(\partial_\mu\bar\psi_v\bar\sigma^\mu
\partial_\mu\psi_v + \partial_\mu\bar\psi_i\bar\sigma^\mu\psi_i
+ \partial_\mu\bar\lambda_A \bar\sigma^\mu\lambda_A\Big)
\nonumber\\
&&-\ \frac{1}{4} F^{\mu\nu}F_{\mu\nu}
- \frac{1}{2}\Big(2g_A^2(M_{\rm GS}^2 + q_i^2|\phi_i|^2)\Big)
A^\mu A_\mu  + J^\mu_A A_\mu\nonumber\\ &&- \
\left( (\sqrt{2}g_A M_{\rm GS}) \psi_v\lambda_A  +
(\sqrt{2} q_i\phi_i^*)\psi_i\lambda_A  + {\rm h.c.}\right)
- \frac{g_A^2}{2}\left(\sqrt{2} M_{\rm GS} t_v - q_i
|\phi_i|^2 \right)^2\nonumber\\&&
 -\  \left(\frac{1}{4}
m_{3/2}\psi^2_v + {\rm h.c.}\right)
+\frac{1}{4} m_{3/2}^2 t_v^2 
+ \left(\frac{\alpha_A}{\delta_{\rm GS}}
m_{3/2}^2\right) \sqrt{2}M_{\rm GS} t_v ,
\eea
where
 $\{ t_v+ i\varphi_v, \psi_v\}$ is the visible
 sector modulus supermultiplet, $t_v$ and $\varphi_v$ are the
 real part and the imaginary part of the scalar modulus respectively,
$\psi_v$ is the fermion component,
$\{ \phi_i, \psi_i\}$ is the chiral matter field  supermultiplet,
$\phi_i$ is the complex scalar, $\psi_i$ is the fermion component,
$\{ A_\mu, \lambda_A\}$ is the $U(1)_A$ gauge supermultiplet
in the WZ gauge, $A_\mu$ is the gauge field and $\lambda_A$ is the
gaugino, the $U(1)_A$ current
$J^\mu_A =\Big( q_i \bar\psi_i\bar\sigma^\mu\psi_i
+ iq_i(\phi_i^*\partial^\mu\phi_i-\phi_i\partial^\mu\phi_i^*)
+ \sqrt{2}M_{\rm GS}\partial^\mu\varphi_v\Big)$, and finally
$M_{\rm GS}$ is the square root of the vacuum value
of ${\cal M}^2_{\rm GS}$ in (\ref{super scale}) :
$M_{\rm GS}= \delta_{\rm GS}t_b^{-3/4}$.

The first three lines of (\ref{treeL}) represent a supersymmetric 
part of the Lagrangian, while the last line is induced by 
the SUSY breaking of the large volume modulus $T_b$. 
Note that the  modulus $t_v$ has 
a soft mass squared of the order $m_{3/2}^2$,
whereas the matter field $\phi_i$ does not have such term.
This difference can be easily understood as follows.
Since the matter field is localized on the MSSM 4-cycle,
the K\"ahler metric of $\Phi_i$ is suppressed by $t_b^{-1}$ 
as in (\ref{treeL}). Thus, the SUSY breaking of $T_b$ is 
not transmitted to $\Phi_i$ and no soft terms are generated at the tree-level.
On the other hand, the K\"ahler metric of $T_v$ is suppressed 
by a inverse power of the Calabi-Yau volume $t_b^{-3/2}$. 
In this case, due to the additional suppression factor $t_b^{-1/2}$, 
sequestering does not work. The resulting SUSY breaking mass squared 
is of the order of $|F^{T_b}/t_b|^2\sim m_{3/2}^2$. 
The SUSY breaking Majorana mass term of 
$\psi_v$ also comes from the K\"ahler potential for the same reason.
The linear term of $t_v$ which is proportional to $\alpha_A$ 
originates from the moduli mixing term,
 $(t_A-\alpha_A\ln t_b)$ in the K\"ahler potential. 

The mass squared of the $U(1)_A$ gauge boson 
$A_\mu$ is given by the supersymmetric contribution, $2 g_A^2(
M_{\rm GS}^2 + q_i^2|\phi_i|^2)$. We want to consider
the case that the gauge boson gets its mass mostly from the
St\"uckelberg mechanism, i.e. $M_{\rm GS}^2 \gg q_i^2|\phi_i|^2  $. 
In this limit, one-loop correction to the scalar potential 
is generated  as follows.
 \bea \label{CW}
 \Delta V_{1-{\rm loop}}(\phi_i) = -\frac{\Lambda^4}
 {128\pi^2}{\rm Str}1
 + \frac{\Lambda^2}{64\pi^2}{\rm Str}M^2 +
 \frac{1}{64\pi^2} {\rm Str}M^4 \left(\ln
 \frac{M^2}{\Lambda^2} -\frac{3}{2}\right),
 \eea
where  $\Lambda$ is the cut-off scale which is independent of $\phi_i$, and
$M^2 = M^2(|\phi_i|^2)$ is the $\phi_i$ dependent tree-level 
mass squared matrix for $A_\mu$, $t_v$, $\phi_i$,  $\psi_v$, $\lambda_A$, 
and $\psi_i$. 
Since the visible sector is localized on the vanishing cycle, 
the natural cut-off scale of the 4D effective field theory is the string scale, 
$\Lambda\sim M_{\rm string}\sim t_b^{-3/4}$. 
The mass of the $U(1)_A$ gauge boson
is of the order of $M_{\rm GS}\sim 
\delta_{\rm GS}t_b^{-3/4}\sim M_{\rm string}/8\pi^2$ 
which is quite below the cut-off scale, so 
we can safely calculate the one-loop correction of (\ref{CW}) 
including all fields discussed above. 
In \cite{Conlon:2009xf}, it was argued that 
in the case of D3 branes at orbifold singularities, 
the cut-off scale is given by the winding scale 
$\Lambda \sim M_{\rm wind}\sim t_b^{-1/2}$ which  
is much bigger than the mass of the $U(1)_A$ gauge boson. 
For all cases, the $U(1)_A$ vector superfield can be included 
in the effective field theory. 
In order to see the cut-off dependence of the soft terms explicitly, 
we do not fix $\Lambda$ as a specific value
during calculation. After calculation, we will  
discuss its effect on the soft terms.

 If we ignore the SUSY breaking terms specified 
in the last line of (\ref{treeL}), the vacuum will be described by 
$D$-flat condition, $\sqrt{2}M_{\rm GS} t_v= q_i |\phi_i|^2$.
Then, a complex scalar field which spans the $D$-flat direction,
and a linear combination of $\psi_v$ and $\psi_i$ which 
does not appear in the third line of (\ref{treeL}) remain massless.
Masses of the other fields are all the same as $2g_A^2(M_{\rm GS}^2 
+ q_i^2|\phi_i|^2)$, and  hence (\ref{CW}) is vanishing. 
Now let us correctly count the SUSY breaking terms.
 By diagonalizing $M^2(|\phi_i|^2)$ and
 expanding the mass eigenvalues in powers of $q_i^2|\phi_i|^2$, we get the following
mass squared spectrum at the leading order.  For bosons,
\bea \label{massB}
A_\mu\ &:& \
M_A^2 = 2g_A^2 \Big( M_{\rm GS}^2 + q_i^2|\phi_i|^2\Big), \nonumber\\
t_v \ &:& \  M_{t_v}^2= M_A^2 -m_{3/2}^2\left(\frac{1}{2}-\frac{q_i^2|\phi_i|^2}{2
M_{\rm GS}^2}\right), \nonumber\\  \phi_i \ &:&\ M_{\phi_i}^2
= -m_{3/2}^2\left(\frac{q_i\alpha_A}{\delta_{\rm GS}}+ 
\frac{q_i^2|\phi_i|^2}{2 M_{\rm GS}^2}\right), 
\eea and for fermions,
\bea\label{massF}  \lambda_A^1\ &:&\ M_{\lambda^1_A}^2= M_A^2
+\frac{1}{8}m_{3/2}^2 +   m_{3/2}M_A\left(\frac{1}{2}-\frac{q_i^2|\phi_i|^2}
{2M_{\rm GS}^2}\right),
\nonumber\\
\lambda_A^2 \ &:&\ M_{\lambda^2_A}^2 = M_A^2
+\frac{1}{8}m_{3/2}^2 -   m_{3/2}M_A\left(\frac{1}{2}-\frac{q_i^2|\phi_i|^2}
{2M_{\rm GS}^2}\right),\nonumber\\
\psi_i \ &:&\ M_{\psi_i}^2 = {\cal O}\left( \frac{m_{3/2}^2|\phi_i|^4 }
{M_{\rm GS}^4}\right),
\eea where $\lambda_A^{1}, \lambda_A^2$ are the
mass eigenstates of the heavy fermions.
In this mass spectrum, the SUSY breaking effect 
induced by the last line of (\ref{treeL})
is reflected on the terms proportional to $m_{3/2}$. 
Although (\ref{massB}) and (\ref{massF}) are 
evaluated directly from (\ref{treeL}), 
one can calculate $M_{\phi_i}^2$ 
from the tree-level effective scalar potential of $\phi_i$
in which $t_v$ is integrated out along the 
$D$-flat direction, $\sqrt{2}M_{\rm GS} t_v\approx q_i|\phi_i|^2$. 
Then, the effective scalar potential $V_{\rm eff}(\phi_i)=
-(q_i\alpha_A/\delta_{\rm GS})m_{3/2}^2 |\phi_i|^2
-(m_{3/2}^2/8M_{\rm GS}^2)q_i^2|\phi_i|^4$
and $M_{\phi_i}^2  =\partial_{\phi_i}\partial_{\phi_i^*}
V_{\rm eff}(\phi_i)=
-m_{3/2}^2\left(q_i\alpha_A/\delta_{\rm GS}+ q_i^2|\phi_i|^2
/2M_{\rm GS}^2\right)$ is obtained. 

It is straightforward to calculate $\Delta V_{1-{\rm loop}}$ by
substituting (\ref{massB}), (\ref{massF}) to (\ref{CW}).
There is no 
soft  mass contribution from  $\Lambda^2{\rm Str}M^2/64\pi^2$. 
However, the last term of the RHS of (\ref{CW}), 
\bea \label{CW1}
&&\frac{1}{64\pi^2}{\rm Str} M^4\ln\frac{M^2}{e^{3/2}\Lambda^2}\nonumber\\
&=& \frac{1}{64\pi^2}\left(3 M_A^4 \ln\frac{M_A^2}{e^{3/2}\Lambda^2}+
M_{t_v}^4\ln\frac{M_{t_v}^2}{e^{3/2}\Lambda^2}
-2M_{\lambda_A^1}^4\ln\frac{M_{\lambda_A^1}^2}{e^{3/2}\Lambda^2}
-2M_{\lambda_A^2}^4\ln\frac{M_{\lambda_A^2}^2}{e^{3/2}\Lambda^2}\right)
\nonumber\\
&&+\ \frac{1}{64\pi^2}\left( M_{\phi_i}^4 \ln\frac{M_{\phi_i}^2}{\Lambda^2}
-2M_{\psi_i}^4\ln\frac{M_{\psi_i}^2}{\Lambda^2}\right),
\eea
should be carefully treated.
Notice that the masses of heavy fields ($A_\mu$, $t_v$, 
and $\lambda_A^{1,2}$) are independent of $\alpha_A$, i.e. 
independent of moduli mixing. 
Thus soft SUSY breaking terms induced by the second line of 
the RHS of (\ref{CW1}) can be called model-independent contribution.
We find that the induced soft mass for $\phi_i$
is also cut-off independent at the one-loop level.  
This is quite reasonable, since it corresponds to the $U(1)_A$ threshold correction.
Suppose $\alpha_A=0$, then there is no soft scalar mass contribution from the third line.
Even if $\alpha_A$ is nonzero, its contribution is suppressed by 
$(m_{3/2}^2/M_{\rm GS}^2)$ compared to the model-independent contribution. 
Accordingly, 
\bea\label{CW2} \Delta V_{1-{\rm loop}}(\phi_i)={\rm constant} -
\frac{g_A^2 q_i^2  }{16\pi^2} m_{3/2}^2|\phi_i|^2
+ {\cal O}\left(\frac{m_{3/2}^2\Lambda^2 |\phi_i|^4}{8\pi^2 M_{\rm GS}^4},
\frac{m_{3/2}^2|\phi_i|^4}{8\pi^2M_{\rm GS}^2},
\frac{\alpha_Am_{3/2}^4 |\phi_i|^2}{\delta_{\rm GS}8\pi^2 M_{\rm GS}^2}\right),\eea
where ``constant" implies that the value is independent of $\phi_i$, 
the scalar mass squared,
\bea\label{MIcomponent}
\Delta m_i^2 = -\frac{g_A^2 q_i^2}{16\pi^2} m_{3/2}^2,
\eea
comes from the second line of the RHS of (\ref{CW1}).
The scalar mass contribution from the last term of (\ref{CW2}) can be 
ignored. The term which depends on the cut-off scale 
is the quartic potential of $|\phi_i|$, so
its effect on the scalar mass is negligible regardless of 
taking $\Lambda$ as $M_{\rm string}$ or $M_{\rm wind}$. 
The value (\ref{MIcomponent}) is identical with
the model-independent contribution of (\ref{soft term2}) obtained by 
setting ${\cal M}_{\rm UV}^2\simeq t_b^{-1}$.

\section{Small moduli $F$-components in the LVS}
\label{append2}

In this appendix, we will exhibit the result of (\ref{smallmod}) explicitly.
We begin from the effective K\"ahler potential (\ref{eff kahler})
and superpotential (\ref{eff super}) constructed by integrating out 
the $U(1)_A$ vector superfield. 
The model consists of a single large volume modulus $T_b$, and
$n_s$ small moduli $T_{sj}$. For each  $T_{sj}$,
there exists non-perturbative correction
to the superpotential ($n_s=n_w$). Also there is no
one-loop induced moduli mixing between $T_{sj}$ and $T_b$ ($\alpha_A=0$).
The matter sector does not have an important role 
for evaluating moduli $F$-terms, so we can ignore it. Then,
the K\"ahler potential and superpotential for the moduli sector are given by
\bea
K_{\rm eff} &=& - 3\ln t_b + \frac{\hat K(\vec t_s) -\xi_{\alpha'}}{t_b^{3/2}}
+ {\cal O}(t_b^{-3}), \nonumber\\
W&=&  W_0 + \sum_j A_a e^{-a_jT_{sj}},
\eea
where
\bea\hat K(\vec t_s)&=&\Delta K(\vec t_s, t^0_A(\vec t_s)) +
\frac{g_A^2 \delta_{\rm GS}^2 }{8\pi^2}
\Delta K''(\vec t_s, t^0_A(\vec t_s))
\ln\frac{2g_A^2\delta_{\rm GS}^2\Delta K''(\vec t_s, t^0_A(\vec t_s))}
{e t_b^{1/2}}\nonumber\\
&= & \Delta K(\vec t_s, t^0_A(\vec t_s))\left(1+
{\cal O}\Big(\frac{1}{(8\pi^2)^3}\Big)\right).\eea
Since the $U(1)_A$ threshold correction is three-loop suppressed,
we set the argument of $\hat K$ just $\vec t_s$ as $\Delta K$.
Due to the no-scaler structure of the tree-level K\"ahler potential,
$\hat K$ has the following scaling behavior
\bea \label{scaling}
\hat K(\lambda \vec t_{s})=\lambda^{3/2} \hat K(\vec t_s)
\eea up to (string) loop corrections. The corresponding scalar potential is
given by  \bea
V_F &=& e^{K}\left( K^{I\bar J} D_I W D_{\bar J} W^* - 3 |W|^2\right)
\nonumber\\
&=&\frac{1}{t_b^{3/2}}\left(\hat K^{i  j}
(\partial_{T_{si}} W)(\partial_{T_{s j}^*}W^*) -\frac{1}{2}
\Big(\hat K^{i j}\hat K_{ j}(\partial_{T_{si}} W)
(W^*t_b^{-3/2})+ h.c.\Big)\right.\nonumber\\
&&\left.\hskip1cm+\ \frac{1}{4}\Big(\hat K^{i j}
\hat K_i\hat K_{ j}-3 \hat K + 3\xi_{\alpha'}\Big)\left|
Wt_b^{-3/2}\right|^2\right)+{\cal O}(t_b^{-6})
\eea
where $\hat K_i=\partial_{t_{si}}\hat K(\vec t_s)$,
$\hat K_{i j}=\partial_{t_{si}}\partial_{t_{sj}}\hat K$,
$\hat K^{i  j}=(\hat K_{i j})^{-1}$.
In leading order of  $1/t_b$ expansion, the stationary condition
$\partial_{T_b}V_F=\partial_{T_{sj}}V_F=0$ gives
\bea \partial_{T_{sk}}V_F=0& : & \Big(\hat K^{i j}
(\partial_{T^*_{sj}}W^*) -\frac{1}{2}\hat K^{i  j}
\hat K_{ j}(W^*t_b^{-3/2})\Big)\Big(\partial_{T_{sk}}
\partial_{T_{si}}W\Big) \nonumber\\ &&+\ \frac{1}{2}
\Big(\hat K^{i  j}_k(\partial_{T_{s j}^*} W^*)
- (\hat K^{i j}_k\hat K_{ j} +\delta^i_k)(W^*t_b^{-3/2}) \Big)
\Big(\partial_{T_{si}}W\Big) + h.c. \nonumber\\ &&+\
\frac{1}{4}\Big(\hat K^{i  j}_k \hat K_i\hat K_{ j}
- \hat K_k\Big)\Big|Wt_b^{-3/2}\Big|^2=0,\nonumber\\
\partial_{t_b}V_F= 0 &: & \hat K^{i  j}(\partial_{T_{si}} W)
(\partial_{T_{sj}^*}W^*)-2\Big(\hat K^{i j}K_{ j}
(\partial_{T_{si}}W)(W^*t_b^{-3/2})+h.c.\Big)
\nonumber\\&&+\ \frac{3}{4}\Big(\hat K^{i j}
\hat K_i\hat K_{ j}-3 \hat K + 3\xi_{\alpha'}\Big)
\Big|Wt_b^{-3/2}\Big|^2=0.\eea In our field basis,
$t_{sk}\partial_{T_{sk}}\partial_{T_{si}}W= -(a_it_{si})
\delta_{ki}\partial_{T_{si}}W$. We would like to 
find the solution in the large volume limit.
Such limit corresponds to $a_j t_{sj} \gg {\cal O}(1)$, and
$|t_{si}\partial_{T_{si}}^2 W| \gg |\partial_{T_{si}} W|$. Then the solution
can be evaluated perturbatively as follows.
\bea\label{sol}
\partial_{T_{si}}W  &=& - a_iA_i  e^{-a_i T_{si}}=\frac{1}{2}
(1-\epsilon_i)\hat K_i (Wt_b^{-3/2}),\nonumber\\
\sum_j\left(\frac{\hat K^{ij}\hat K_j}{t_i}\right)\epsilon_j
&=& \frac{1}{a_i t_{si}}\left(2-\sum_{j,k} \frac{\hat K_i^{jk}
\hat K_j\hat K_k}{2\hat K_i}\right)
\Big(1+{\cal O}(\epsilon_i)\Big), \nonumber\\
\xi_{\alpha'}&=& \hat K\left(1 + \sum_{ij}
\left(\frac{2 \hat K^{ij}\hat K_i\hat K_j }{9\hat K}\right)\epsilon_i
+ {\cal O}(\epsilon_i^2)\right).
\eea Notice that there is no sum for an index $i$.
We assume that  $A_i$ and the vacuum value of $\hat K_i$
 are of order one. However,
the gravitino mass $m_{3/2}=e^{K/2} W= (W_0 t_b^{-3/2})
(1+{\cal O}(t_b^{-3/2}))$ would be around ${\rm TeV}$
so that it is  hierarchically much smaller than one.
From the first equation of (\ref{sol}),
\bea
a_i t_{si} &=&2\ln\frac{M_{\rm Pl}}{|m_{3/2}|}-
2\ln\frac{\hat K_i(1-\epsilon_i)}{2|a_i A_i|}=
2\ln\frac{M_{\rm Pl}}{|m_{3/2}|}+ {\cal O}(1).
\eea
$a_it_{si}$ are universal at the leading order.
On the one hand, due to the scaling behavior of $\hat K$ given by (\ref{scaling}),
it is easily identified that $\sum_{ij}\hat K^{ij}_k
\hat K_i \hat K_j = \hat K_k$. Then,
the $F$-terms of the small moduli can be obtained as  \bea
\frac{F^{T_{si}}}{t_{si}}&=& -\frac{1}{t_{si}}e^{K/2}K^{T_{si} \bar J} D_{\bar J}W^*=
\sum_j\left(\frac{\hat K^{ij}\hat K_j}{t_{si}}\right) \epsilon_j^* (W^* t_b^{-3/2})
\nonumber \\ &=& \frac{m_{3/2}^*}{\ln|M_{Pl}/m_{3/2}|}
\left(\frac{3}{4}+ {\cal O}\left(\frac{1}{\ln|M_{\rm Pl}/m_{3/2}|},
\frac{\hat K_{\rm loop}}{\hat K}\right) \right),
\eea where $\hat K_{\rm loop}$ stands for the perturbative
correction to $\hat K$ which breaks the no-scale form of K\"ahler potential.


\end{document}